\documentclass[fleqn,usenatbib]{mnras}
\usepackage{newtxtext,newtxmath}
\usepackage[T1]{fontenc}
\DeclareRobustCommand{\VAN}[3]{#2}
\let\VANthebibliography\thebibliography
\def\thebibliography{\DeclareRobustCommand{\VAN}[3]{##3}\VANthebibliography}
\usepackage{graphicx}	
\usepackage{amsmath}	
\usepackage{mathtools}
\usepackage{placeins}
\usepackage{bm}
\usepackage{afterpage}
\usepackage{hyperref}
\usepackage{float}
\floatstyle{plaintop}
\restylefloat{table}
\usepackage[labelfont=bf]{caption}
\usepackage{layouts}
\usepackage{float}

\begin{document}

\title[Propagating $m$-bias to parameter estimation]{Propagating spatially-varying multiplicative shear bias to cosmological parameter estimation for stage-IV weak-lensing surveys}
\author[C. Cragg et al.]{
Casey Cragg,$^{1}$\thanks{E-mail: casey.cragg@physics.ox.ac.uk}
Christopher A. J. Duncan,$^{1}$
Lance Miller$^{1}$
and David Alonso$^{1}$
\\
$^{1}$Department of Physics, University of Oxford, Denys Wilkinson Building, Keble Road, Oxford OX1 3RH, UK
}
\date{Accepted XXX. Received YYY; in original form ZZZ}

\label{firstpage}
\pagerange{\pageref{firstpage}--\pageref{lastpage}}
\maketitle

\begin{abstract}
We consider the bias introduced by a spatially-varying multiplicative shear bias ($m$-bias) on tomographic cosmic shear angular power spectra. To compute the bias in the power spectra, we estimate the mode-coupling matrix associated with an $m$-bias map using a computationally-efficient pseudo-$C_{\ell}$ method. This allows us to consider the effect of the $m$-bias to high $\ell$. We then conduct a Fisher matrix analysis to forecast resulting biases in cosmological parameters. For a \textit{Euclid}-like survey with a spatially-varying $m$-bias, with zero mean and rms of 0.01, we find that parameter biases reach a maximum of $\sim 10 \%$ of the expected statistical error, if multipoles up to $\ell_{\mathrm{max}}=5000$ are included. We conclude that the effect of the spatially-varying $m$-bias may be a sub-dominant but potentially non-negligible contribution to the error budget in forthcoming weak lensing surveys. We also investigate the dependence of parameter biases on the amplitude and angular scale of spatial variations of the $m$-bias field, and conclude that requirements should be placed on the rms of spatial variations of the $m$-bias, in addition to any requirement on the mean value. We find that, for a \textit{Euclid}-like survey, biases generally exceed $\sim 30 \%$ of the statistical error for $m$-bias rms $\sim 0.02 - 0.03$ and can exceed the statistical error for rms $\sim 0.04 - 0.05$. This allows requirements to be set on the permissible amplitude of spatial variations of the $m$-bias that will arise due to systematics in forthcoming weak lensing measurements.
\end{abstract}

\bigskip
\section{Introduction}
\label{intro}
A key area of study in cosmology is that of the statistics of the large-scale matter distribution, which can provide constraints on the evolution of cosmic structures, the expansion history of the Universe, the nature and behaviour of dark matter and dark energy, the conditions in the very early Universe, and alternatives to the $\Lambda$CDM cosmological model. One promising probe of the large-scale structure is the phenomenon of weak gravitational lensing, in which the images of distant galaxies are distorted due to the gravitational perturbation of the paths of light rays by matter between the source and observer \citep{jain:1997, kilbingerreview:2015}. Current and future weak lensing surveys aim to measure the spin-2 \textit{cosmic shear} $\bm{\gamma}$, estimated by the ellipticities of source galaxy images, into which a spatially-coherent signal is introduced due to weak lensing. By studying the cosmic shear we can make inferences about the large-scale structure.
\par
Cosmic shear measurements have already been made by previous and ongoing cosmological surveys, including CFHTLenS \citep{cfhtlens:2012}, KiDS \citep{kids:2021} and DES \citep{des:2021}, and shear measurement is one of the main science goals of the forthcoming stage IV surveys, including the \textit{Rubin} Observatory's Legacy Survey of Space and Time (LSST) \citep{lsst:2009}, \textit{Euclid} \citep{euclid:2011} and the \textit{Roman} telescope \citep{roman:2012}. Cosmic shear is a very subtle effect, with distortions introduced of the order of a per cent of the intrinsic galaxy ellipticities \citep{kilbingerreview:2015}. As additional data is contributed by each successive survey, the resulting reduction in statistical uncertainty means that systematic errors need to be more tightly controlled. While previous weak lensing surveys have accounted for systematic effects, forthcoming surveys will require understanding and mitigation of systematics to an unprecedented level of precision, including effects which have not previously been considered. One significant effect of astrophysical and instrumental systematics is the introduction of biases into the measurement of the cosmic shear inferred from galaxy shapes. This can be expressed with a linear bias model:
\begin{equation}
\widetilde{\bm{\gamma}} = \bm{\gamma}(1+m)+c \ , \label{eq:1}
\end{equation}
with $\widetilde{\bm{\gamma}}$ the biased shear field, $m$ the multiplicative “$m$-bias” and $c$ the additive “$c$-bias”. These biases can arise from a number of systematic effects, including galaxy shape measurement errors, selection effects, as well as point spread function (PSF) estimation errors and other instrumental effects \citep{huterer:2006, kilbingerreview:2015, mandelbaum:2018, taylor:2018, pujol:2020}. Different systematic effects can separately introduce additive and multiplicative biases; for example, a $c$-bias can arise from anisotropy in the modelled PSF, while an $m$-bias can result from PSF size errors \citep{huterer:2006}. While it is desirable to estimate and correct for these biases theoretically (eg. from PSF modelling), it may also be possible to (at least partially) detect the $c$-bias empirically, eg. using null tests \citep{uitert:2016}. However, the latter is not possible for the $m$-bias, which must be predicted, eg. from image simulations \citep{kannawadi:2019, pujol:2019}, or inferred from self-calibration during the measurement process \citep{huff:2017, sheldon:2017, sheldon:2020}. For forthcoming stage-IV surveys, it is expected that the shape biases will need to be constrained to the level of $c \lesssim 10^{-4}$, $m \lesssim 2 \times 10^{-3}$ in order to obtain the required sensitivity \citep{massey:2013}.
\par
In this paper, we investigate the effect of a spatially-varying multiplicative bias in the shear measurement on cosmological parameters estimated from the cosmic shear angular power spectrum, $C_{\ell}^{\gamma \gamma}$. In forthcoming surveys we will be concerned with residual shear biases, i.e. biases arising due to imperfect calibration of systematics as a result of uncertainties in systematics models. Such a bias could arise e.g. when there are position-dependent errors in the PSF model. In particular, multiplicative bias is expected to occur due to errors in the assumed size of the PSF \citep{paulin-henriksson:2008}. In optical weak-lensing surveys the typical imaging field size is $\lesssim 1^{\circ}$, corresponding to multipole $\ell \gtrsim 200$, which provides a natural characteristic scale on which the $m$-bias may vary \citep{paykari:2020}. Additionally, there are expected to be PSF variations, and hence the possibility of PSF modelling errors, down to arcminute scales ($\ell \sim 10^{3}$ $-$ $10^{4}$), motivating us to study the effect of a spatially-varying $m$-bias on a wide range of scales. There may also be spatially-varying $m$-bias arising from other effects in the shear measurement, such as may arise from variations in galaxy density or morphology. 
\par
Previous work has been conducted in this area, in particular by \cite{kitching:2019}, \cite{kitching:2020}, in which it was found that a spatially-varying $m$-bias has a negligible effect on the shear power spectrum compared with the mean of the $m$-bias field, and should be negligible also when the $m$-bias field averages to zero across the sky. However, \cite{kitching:2019} did not extend the fully non-linear analytic calculation beyond a maximum multipole of $\ell_{\mathrm{max}} =$ 32 (64 in \cite{kitching:2020}). In \cite{kitching:2020}, a linear approximation of the bias in $C_{\ell}^{\gamma \gamma}$ was also calculated up to $\ell_{\mathrm{max}} = 2048$ and compared with a forward model, with differences found that were $\sim 4$ orders of magnitude smaller than the cosmic variance. \cite{kitching:2020} also calculated the bias in the estimated amplitude of the shear power spectrum resulting from marginalisation over a prior on the mean $m$, and placed requirements on the bias and variance of that distribution with respect to the true mean $m$. That work did not consider the parameter bias arising due to the convolutive effect on the shear $C_{\ell}$ arising due to the spatial variations of the $m$-bias field.
\par
\cite{kitching:2019}, \cite{kitching:2020} propagated shear biases into biases in the $C_{\ell}^{\gamma \gamma}$ using a fully general calculation which allowed for an $m$-bias field which is a complex-valued spin field (i.e. the m-bias field was allowed to have a rotation), resulting in a rapid scaling of the computational time required as $\ell_{\mathrm{max}}^{5}$. We show that the resulting stringent upper limit on the range of multipoles that can be included in the analysis is insufficient, because a spatially-varying $m$-bias with some characteristic scale (which may be large or small) will bias the $\smash{C_{\ell}^{\gamma\gamma}}$ down to small scales due to mode-mode coupling. In this work, we restrict our analysis to consider a real-valued spin-0 $m$-bias, as is commonly assumed in shear analyses and is expected to result from realistic shear systematics; the real part was predicted to be dominant compared to the imaginary part of the $m$-bias field by \cite{kitching:2019}. We apply a computationally efficient pseudo-$C_{\ell}$ formalism taking advantage of the resulting symmetries to determine the effect on the shear power spectrum of a spatially-varying $m$-bias. Exploiting the assumption of a spin-0 $m$-bias field, this method scales as $\ell_{\mathrm{max}}^{3}$ \citep{alonso:2019}. This allows us to compute the bias in the $C_{\ell}^{\gamma \gamma}$ down to scales corresponding to $\ell_{\mathrm{max}} \sim 10^{3}-10^{4}$. The calculation to high $\ell$ is tractable because, in the case of a spin-0 $m$-bias, orthogonality of the Wigner 3-j symbols involved in the calculation of the mode-coupling matrix reduces the dimensionality of the calculation. We consider the effect of a random $m$-bias map generated according to an angular power spectrum with some characteristic scale defined by a peak multipole $\ell_{\mathrm{peak}}$ and width $\sigma_{m}$. In our analysis we consider the bias in the $\smash{C_{\ell}^{\gamma\gamma}}$ arising due to a range of values for $\ell_{\mathrm{peak}}$ and $\sigma_{m}$ and subsequently utilise a Fisher-matrix analysis to predict the sensitivity of cosmological parameters to the spatially-varying $m$-bias for a typical next-generation survey. While previous works have propagated shear biases into biases in cosmological parameters using a Fisher matrix analysis, these have either ignored the spatial variations of the $m$-bias (eg. \cite{amara:2008}), relied on assumptions about the $\ell$-coupling of the $C_{\ell}$s (eg. \cite{massey:2013,kitching:2016}), or did not carry out the full propagation of the shear bias from angular space (eg. \cite{taylor:2018}, which propagated the shear bias power spectra into dark energy figures of merit); this work is the first to calculate the full mode-coupling matrix corresponding to $m$-bias maps and propagate into biases in cosmological parameters. We also compute the $BB$ power spectra which result from the mode-mixing effect arising due to the spatially-varying $m$-bias and compare it to the bias in the $EE$ power spectra. Note that we do not consider the additive shear bias in the present work, as its form is expected to be strongly dependent on the particular systematics from which it results, whereas here we consider a generic form for the $m$-bias which may be broadly relevant to future surveys. We are primarily interested in the effect of the $m$-bias at high multipoles in the $C_{\ell}^{\gamma \gamma}$, and defer investigation of the $c$-bias to future work.
\par
This paper is organised as follows. In section 2 we describe our methods, including the generation of our fiducial $C_{\ell}^{\gamma \gamma}$s, $m$-bias maps, the pseudo-$C_{\ell}$ formalism used to calculate the biased $C_{\ell}^{\gamma \gamma}$s, and the Fisher-matrix analysis used to forecast resulting biases in cosmological parameters. In section 3 we show our results, including the residual $C_{\ell}^{\gamma \gamma}$s, and biases and the ratios of biases to the respective $1\sigma$ uncertainty in cosmological parameter estimates. In section 4 we discuss the results of our analysis, and in section 5 we summarise the findings of our work.
\section{Methods}
\label{methods}
\subsection{Pseudo-$C_{\ell}$ formalism}
\label{pseudo}
Consider a spin-$s_{a}$ vector field $\bm{a}(\bm{\Omega})$, where $\bm{\Omega} = (\theta,\phi)$ is the angular unit vector on the sphere, which is observed with a mask (or set of weights) $W_{a}(\bm{\Omega})$, such that the observed field $\widetilde{\bm{a}}(\bm{\Omega}) = \bm{a}(\bm{\Omega}) W_{a}(\bm{\Omega})$. We may calculate the spherical harmonic expansion coefficients of the masked field as follows \citep{hivon:2002,kogut:2003}:
\begin{align}
\widetilde{\bm{a}}_{\ell m} &= \int_{\bm{\Omega}}d\bm{\Omega}\mathsf{Y}_{\ell m}^{s_{a} \dag}(\bm{\Omega})\bm{a}(\bm{\Omega})W_{a}(\bm{\Omega}) \label{eq:2} \\
&= \sum_{\ell ’ m’ }\int_{\bm{\Omega}}d\bm{\Omega}W_{a}(\bm{\Omega})\left(\mathsf{Y}_{\ell m}^{s_{a} \dag}(\bm{\Omega})\cdotp\mathsf{Y}_{\ell ’ m’}^{s_{a}}(\bm{\Omega})\right)\cdotp\bm{a}_{\ell ’ m’} \label{eq:3} \\ 
&\equiv \sum_{\ell ’ m’}\mathsf{K}_{\ell m \ell ’ m’}^{s_{a}}[W_{a}]\cdotp\bm{a}_{\ell ’ m’} \ , \label{eq:4}
\end{align}
where $\mathsf{Y}_{\ell m}^{s_{a}}$ is a $2 \times 2$ matrix defined in terms of the spin-$s_{a}$-weighted spherical harmonic of degree $\ell$ and order $m$ \citep{zaldarriaga:1997,alonso:2019}, and superscript $\dag$ here denotes conjugate transposition; and $\mathsf{K}_{\ell m \ell ’ m’}^{s_{a}}[W]$ is the spin-$s_{a}$ mode-mixing kernel describing the mode-mode coupling of a spin-$s_{a}$ field due to $W_{a}$, and is also a $2 \times 2$ matrix. If we also consider a spin-$s_{b}$ vector field $\bm{b}(\bm{\Omega})$ observed with mask $W_{b}(\bm{\Omega})$, then the $2 \times 2$ matrix of angular power spectra of vector fields $\widetilde{\bm{a}}(\bm{\Omega})$ and $\widetilde{\bm{b}}(\bm{\Omega})$ is estimated according to the following \textit{pseudo-}$C_{\ell}$ formalism \citep{hivon:2002,kogut:2003,alonso:2019}:
\begin{align}
\mathrm{vec}&\left[\langle \widetilde{\mathsf{C}}_{\ell}^{\widetilde{a}\widetilde{b}} \rangle\right] = \frac{1}{2\ell+1} \nonumber \\
&\times \sum_{\ell'} \sum_{m m'} \left(\mathrm{vec}\left[\mathsf{K}_{\ell m \ell' m'}^{s_{a}}[W_{a}]\right]\cdotp \mathrm{vec}\left[\mathsf{K}_{\ell m \ell' m'}^{s_{b}}[W_{b}]\right]^{\dag}\right) \nonumber \\
& \; \cdotp  \mathrm{vec}\left[\langle \mathsf{C}_{\ell'}^{ab} \rangle\right] \label{eq:5} 
\\
&\equiv \sum_{\ell'} \mathsf{M}_{\ell \ell'}^{s_{a}s_{b}} \mathrm{vec}\left[\langle \mathsf{C}_{\ell'}^{ab} \rangle\right] \ , \label{eq:6} 
\end{align}
where the vectorisation operation $\mathrm{vec}[\mathsf{x}]$ creates an $ab \times 1$ vector from an $a \times b$ matrix $\mathsf{x}$ by transposing rows and concatenating them into a single column vector \citep{hamimeche:2008}. Note that the angle brackets denote an ensemble average; we will henceforth omit these for brevity. $\mathsf{M}_{\ell\ell’}^{s_{a}s_{b}}$ is the \textit{mode-coupling matrix} which convolves the true power spectra $\mathsf{C}_{\ell’}^{ab}$ with the cross power spectrum of the masks, and is calculated from the spin-weighted spherical harmonic decompositions of $W_{a}(\bm{\Omega})$ and $W_{b}(\bm{\Omega})$. Considering equation $(1)$, we can consider $1 + m(\bm{\Omega})$ as a non-binary mask that is applied to the true shear field $\bm{\gamma}(\bm{\Omega})$, and can therefore calculate the pseudo-$C_{\ell}$ of the biased shear field using equation $(6)$. Note that for a spin-$2$ field such as the cosmic shear, the mode-coupling matrix also mixes $E$- and $B$-modes, generating $BB$-power which is otherwise expected to be subdominant in the intrinsic shear field.
\par
As equation \eqref{eq:5} includes a sum over three indices of a product that is integrated over the whole sphere, calculation of the mode-coupling matrix is computationally expensive in the general case. However, it is possible to express the mode-mixing kernels in terms of Wigner 3-$j$ symbols \citep{edmonds:1957,hivon:2002,kogut:2003}. In the case that the $m$-bias is isotropic such that it affects both components of the shear equally, as we consider here and is likely to be the case in real surveys, the Wigner 3-$j$ symbols obey an orthogonality relation that allows us to collapse the summation in indices $m_{1},m_{2}$, greatly simplifying the calculation to a sum over the angular cross-power spectrum of the masks weighted by 3-$j$ symbols \citep{kogut:2003}. \texttt{NaMaster} employs this method (and additionally replaces in the calculation the $C_{\ell}$ of the masks with their pseudo-$C_{\ell}$, which is more computationally efficient to calculate) \citep{alonso:2019}. This allows for a significant reduction in the computational cost of the calculation of the angular power spectra of masked fields, reducing the scaling of the calculation by a factor of $\ell_{\mathrm{max}}^{2}$, thereby allowing calculation beyond the maximum multipoles that would be of interest for stage IV weak-lensing experiments, an improvement (subject to our assumption of spin-0 $m$) on the scaling found in the general calculation by \cite{kitching:2019}.
\par
The mode-coupling matrix $\mathsf{M}_{\ell\ell’}^{s_{a}s_{b}}$ depends only on the power spectrum of the mask \citep{alonso:2019}. As the clustering of the $m$-bias map considered here is by construction fully described by its angular power spectrum, it is not necessary to consider multiple realisations of the $m$-bias map for each of our values of $\ell_{\mathrm{peak}}$, $\sigma_{m}$ and $m$-bias rms. It should be noted however that \texttt{NaMaster} takes as its input the mask and computes an estimate of the mask $C_{\ell}$, which is used to compute the mode-coupling matrix. In our case, this mask is a noisy realisation of the $m$-bias map, which is a random field. This results in a noisy estimate of the $m$-bias map $C_{\ell}$ which is used to calculate the $\mathsf{M}_{\ell\ell’}^{s_{a}s_{b}}$. While this results in a loss of optimality of the $\mathsf{M}_{\ell\ell’}^{s_{a}s_{b}}$ estimate, this is not expected to be a significant effect when the smallest scales $\gtrsim 10^{4}$ are neglected, and so we did not modify \texttt{NaMaster} to take the $m$-bias map $C_{\ell}$ as its input directly in order to avoid this effect. This modification would be trivial in a practical application. Additionally, the $m$-bias $C_{\ell}$ will not be known in a real survey, and in a realistic case propagation from angular space (rather than $\ell$-space) will be necessary; this work is also intended to serve as a proof of concept that the full propagation from angular space to $m$-bias maps to cosmological parameter biases is tractable.
\par
In the case of a spatially varying multiplicative bias as considered here, the biased shear field is:
\begin{equation}
\widetilde{\bm{\gamma}}(\bm{\Omega}) = \bm{\gamma}(\bm{\Omega})(1+m(\bm{\Omega})) \ . \label{eq:7}
\end{equation}
We will henceforth refer to $1+m(\bm{\Omega})$ as the $m$\textit{-bias map}. We can therefore apply a pseudo-$C_{\ell}$ approach to describe the effect of a spatially-varying $m$-bias on the shear $C_{\ell}$s. This will allow us to compute the bias in the cosmic shear angular power spectra due to $m({\bf{\Omega}})$ with a weaker scaling with $\ell_{\mathrm{max}}$ than was obtained in previous work in this area (eg. \cite{kitching:2019,kitching:2020}), enabling us to probe the impact of the spatially-varying $m$-bias out to much higher multipoles.
\par
Note that here we have neglected both the $c$-bias and any separate spatially constant $m$-bias. While $m(\bm{\Omega})$ could in general contain a spatially constant term, we are seeking specifically to investigate the impact of the spatial variations which have previously been claimed to have negligible effect. Moreover, since the shear is a spin-2 vector field, in general it can be decomposed into $E$- and $B$-modes, and there are therefore in principle four power spectra, $C_{\ell}^{EE}$, $C_{\ell}^{BB}$, $C_{\ell}^{BE}$ and $C_{\ell}^{EB}$; however, in the absence of certain systematic effects (including intrinsic alignments, selection effects and some PSF correction effects), the $B$-mode field is expected to be highly sub-dominant \citep{kilbingerreview:2015} (though not vanishing, as source clustering is expected to introduce some $B$-modes \citep{schneider:2002}). We will therefore not consider any intrinsic $B$-mode field. In our analysis, we calculate the biased $EE$ power spectrum as well as the $BB$ power spectrum generated by the mode-coupling arising due to $m(\bm{\Omega})$. No $E$-$B$ cross-correlation is expected to be generated in the presence of a spin-0 $m$-bias because the parity-odd $EB$, $BE$ power spectra are not mixed with the parity even $EE$, $BB$ power spectra \citep{alonso:2019}. As the shear field is expected to be approximately $B$-mode-free, any observed $B$-mode power may be used to detect the presence of and potentially calibrate residual systematic effects \citep{asgari:2020}. In particular, the $BB$ power spectra contain information about the $m$-bias field which may in principle be used for self-calibration \citep{kitching:2020}. However, we do not consider this in our analysis due to uncertainty in the viability of self-calibration of the $m$-bias using the $BB$ power, arising due to potential degeneracies in the $B$-modes generated by different systematics, according to our systematics-agnostic approach.
\par
We do not consider any redshift-dependence of the $m$-bias, and apply the same $m$-bias to all of the tomographic shear power spectra we consider. In a real survey there will be a redshift-dependence of the $m$-bias, eg. due to the decrease in observed galaxy size with redshift; as we here consider a generic form for the $m$-bias, without reference to a specific bias model, we defer consideration of this effect to future work. We also conduct a full-sky analysis and so do not consider the mode-coupling effect due to any survey mask boundary.
\par
We consider a randomly-drawn $m$-bias map which is distributed according to an angular power spectrum which has a Gaussian profile as a function of $\ell$. We choose a certain peak multipole for the $m$-bias $C_{\ell}$, $\ell_{\mathrm{peak}}$, and a standard deviation, $\sigma_{m}$. The $m$-bias $C_{\ell}$ is normalised such that the resulting $m$-bias map has r.m.s. value equal to 0.01, a nominal fiducial value which is a reasonable expectation for a forthcoming stage-IV survey such as \textit{Euclid}. We also investigate how our results vary with this value in order to allow requirements to be set on the amplitude of spatial variations of the $m$-bias. We consider $m$-bias $C_{\ell}$s with $\ell_{\mathrm{peak}}$ in the range $50 - 5000$ and $\sigma_{m}$ in the range $64 - 2048$.
\subsection{Survey model}
\label{survey}
For the fiducial shear power spectra, we assume a redshift distribution of galaxies $n(z)$ of the following form \citep{smail:1994}:
\begin{equation}
n(z) \propto \bigg(\frac{z}{z_{0}}\bigg)^{2}\exp\Bigg[-\bigg(\frac{z}{z_{0}}\bigg)^{\frac{3}{2}}\Bigg] \ , \label{eq:8}
\end{equation}
which is normalised to unity, and where $z_{0} = z_{m}/\sqrt{2}$ with $z_{m}$ the median redshift of the distribution \citep{euclidprep7:2020}. We divide this distribution into 10 equipopulated redshift bins up to a maximum redshift of 3, and convolve the distribution in each bin $n_{i}(z)$ with a probability density function with two slightly offset Gaussian terms, given by \citet{euclidprep7:2020}, to account for the effect of photometric redshift errors with catastrophic outliers. The resulting $n_{i}(z)$ are shown in fig. 1. 
\begin{figure}
\includegraphics[width=0.5\textwidth]{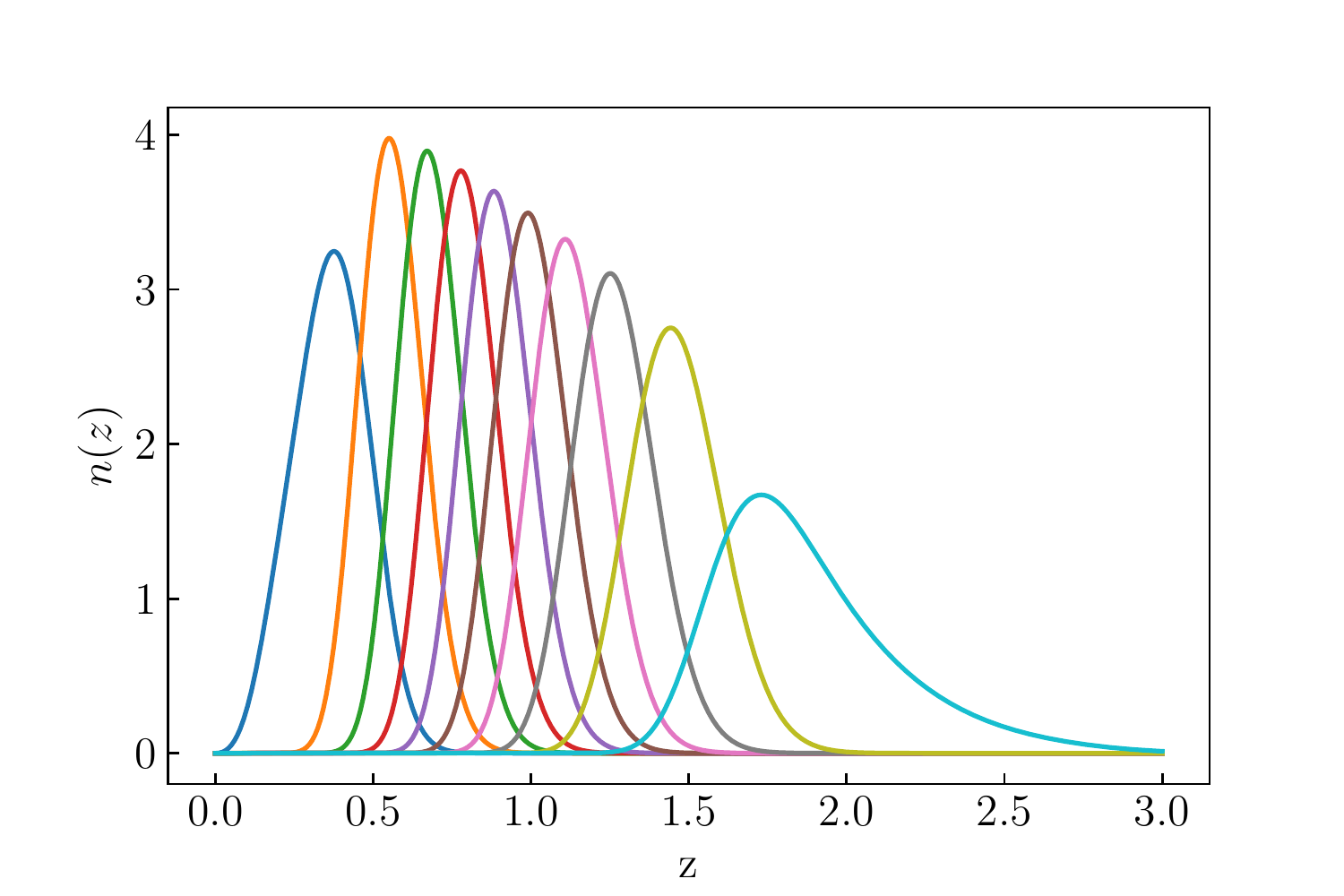}
\caption{Gaussian-convolved $n(z)$ for the 10 tomographic redshift bins.}
\end{figure}
We calculate fiducial shear auto- and cross- ($EE$) power spectra for each pair of bins using \texttt{CCL} \citep{CCL:2019}, denoted $\smash{C_{\ell}^{\gamma_{i}\gamma_{j}}}$ for bin pair $(i,j)$. We assume spatially-flat $\Lambda$CDM with the \textit{Planck} 2018 best-fit cosmological parameter values as our fiducial values for cosmological parameters (see table 1, \citet{planck:2018}). We do not include massive neutrinos in our fiducial cosmological model. In our Fisher-matrix analysis, we consider dark energy with a dynamical equation of state in which the parameters $w_{0}$ and $w_{a}$ are allowed to vary, with fiducial values of $-1$ and $0$ respectively.
\par
\begin{table}
\centering
\caption{Fiducical $\Lambda$CDM cosmological parameter values - cold dark matter density parameter $\Omega_{c}$; baryonic matter density parameter $\Omega_{b}$; reduced Hubble constant $h$; amplitude of the matter power spectrum $\sigma_{8}$; scalar spectra index $n_{s}$ (values from \citet{planck:2018}, table 2, final column); dark energy equation-of-state parameters $w_{0}$ and $w_{a}$.}
\begin{tabular}{| c || c |}
\hline\hline
Cosmological parameter & Fiducial value  \\
\hline\hline
$\Omega_{c}$ &  $0.2607$  \\
$\Omega_{b}$ &  $0.0490$  \\
$h$ &  $0.6766$  \\
$\sigma_{8}$ & $0.810$  \\
$n_{s}$ & $0.9665$  \\
$w_{0}$ & $-1$ \\
$w_{a}$ & $0$ \\
\hline\hline
\end{tabular}
\end{table}
To each of the auto-power spectra $C_{\ell}^{\gamma_{i}\gamma_{i}}$ we add a constant shape-noise term $n$, given by \citep{euclidprep7:2020} for the case of equipopulated redshift bins:
\begin{equation}
n = \frac{\sigma_{e}^{2}N_{z}}{n_{\mathrm{gal}}} \ , \label{eq:9}
\end{equation}
where $\sigma_{e}^{2}$ is the intrinsic ellipticity variance of source galaxies, $N_{z}$ is the number of tomographic redshift bins and $n_{\mathrm{gal}}$ is the sky number density of source galaxies; following \cite{euclidprep7:2020}, we take these to be $\sigma_{e}^{2} = 0.3^{2} = 0.09$,  $N_{z} = 10$ and $n_{\mathrm{gal}} = 30 \  \mathrm{arcmin}^{-2}$.
 For each $m$-bias map under study, we use \texttt{NaMaster} to compute the mode-coupling matrix $\mathsf{M}_{\ell\ell’}^{22}$ and couple this with each $\smash{C_{\ell}^{\gamma_{i}\gamma_{j}}}$ to obtain biased shear power spectra $\widetilde{C}_{\ell}^{\widetilde{\gamma}_{i}\widetilde{\gamma}_{j}}$. We then calculate \textit{residual $C_{\ell}$s}:
\begin{equation}
\Delta C_{\ell}^{\gamma_{i}\gamma_{j}} = C_{\ell}^{\gamma_{i}\gamma_{j}}-\widetilde{C}_{\ell}^{\widetilde{\gamma}_{i}\widetilde{\gamma}_{j}} \ . \label{eq:10}
\end{equation}
Residual shear $C_{\ell}$s are computed using \texttt{NaMaster} up to $\ell_{\mathrm{cut}} = 12288$ (this is set by the resolution of the input \texttt{Healpix} map, where we choose $N_{\mathrm{side}} = \ell_{\mathrm{cut}}/3 = 4096$). We calculate the $\smash{\Delta C_{\ell}^{\gamma\gamma}}$ up to higher multipoles than we consider in our subsequent Fisher matrix analysis in order to obtain more accurate results for the mode-mixing at the scales of interest.
\subsection{Fisher matrix analysis}
\label{fisher}
We conduct a Fisher matrix analysis to forecast biases in cosmological parameters propagated from the bias in the shear $C_{\ell}$s due to the spatially-varying $m$-bias. Given a data vector $\bm{D}_{\ell}$ with non-zero mean consisting of angular power spectra at a particular $\ell$ and parameter-independent data covariance matrix $\mathsf{Cov}(\ell)$,  the Fisher matrix can be approximated as \citep{tegmark:1997,duncan:2014}:
\begin{equation}
F_{\eta\tau} = \sum_{\ell}(\partial_{\eta}\bm{D}_{\ell})(\mathsf{Cov}^{-1})(\ell)\partial_{\tau}\bm{D}_{\ell} \ , \label{eq:11}
\end{equation}
where $F_{\eta\tau}$ is the Fisher matrix and $\partial_{\tau}$ denotes partial differentiation with respect to parameter $\theta_{\tau}$. We take for our data vector at each $\ell$ the vector composed of our fiducial $\smash{C_{\ell}^{\gamma_{i}\gamma_{j}}}$s, which we denote as $\bm{C}_{\ell}$ for brevity; i.e. $\bm{C}_{\ell} = \{C_{\ell}^{\gamma_{0}\gamma_{0}}, C_{\ell}^{\gamma_{0}\gamma_{1}}, \dotsb , C_{\ell}^{\gamma_{(N_{z}-1)}\gamma_{(N_{z}-1)}}\}$, where the redshift bin indices run from 0 to 9, and $\mathsf{Cov}(\ell)$ is the $55 \times 55$ covariance matrix of these power spectra at multipole $\ell$, i.e. a submatrix of the $55N_{\ell} \times 55N_{\ell}$ full block-diagonal covariance matrix, with $N_{\ell}$ the number of $\ell$-modes considered.
\par
The derivatives of the data vector $\partial_{\tau}\bm{C}_{\ell}$ are approximated with a simple two-step derivative:
\begin{equation}
\partial_{\tau}\bm{C}_{\ell} \approx \frac{\bm{C}_{\ell}(\theta_{\tau,\mathrm{fid}} + \delta \theta_{\tau}) - \bm{C}_{\ell}(\theta_{\tau, \mathrm{fid}} - \delta \theta_{\tau})}{2\delta \theta_{\tau}} \ , \label{eq:12}
\end{equation}
where $\theta_{\tau, \mathrm{fid}}$ is the fiducial value of parameter $\theta_{\tau}$ and $\delta \theta_{\tau}$ is a small derivative step in $\theta_{\tau}$. We conduct our analysis with two sets of cosmological parameters for which cosmic shear is expected to be sensitive. In our base parameter set, we consider the cold-dark-matter density parameter, $\Omega_{c}$, that is the ratio of the density of cold dark matter to that of the critical density at the present epoch; $\sigma_{8}$, the amplitude of the linear matter power spectrum on scales of 8$h^{-1}\mathrm{Mpc}$; and the constant dark energy equation-of-state parameter, $w_{0}$. In our extended parameter set, we consider the base parameter set in addition to a parameter capturing the redshift-variation of the dark energy equation-of-state, $w_{a}$, where the evolution of the equation of state parameter is parametrised by $w(z) = w_{0} + w_{a}z/(1+z)$. We take for our derivative steps in these parameters $\{\delta \Omega_{c} = 0.01, \delta \sigma_{8} = 0.005, \delta w_{0} = 0.055, \delta w_{a} = 0.058 \}$. Numerical stability of the derivatives was ensured by selecting values of the derivative steps such that all the Fisher matrix elements were stable with respect to perturbations in the region of the step choice. We consider these parameters to be "free" in our analysis, considering the impact on their inference from the $\smash{C_{\ell}^{\gamma\gamma}}$ due to bias in the shear power spectra, whereas the other parameters listed in table $(1)$ ($h$, $\Omega_{b}$, $n_{s}$) are assumed to be well-known and fixed to their fiducial values.
In our results, we present the bias in $\Omega_{m}$, the total matter density parameter, which in $\Lambda$CDM is the sum of $\Omega_{c}$ and $\Omega_{b}$ (the density parameter of baryonic matter); since we only vary $\Omega_{c}$ and fix the value of $\Omega_{b}$ to the \citet{planck:2018} value, the Fisher matrix elements and the resulting biases that are obtained by varying $\Omega_{c}$ are also the correct results for $\Omega_{m}$ .
\par
The covariance between a pair of shear power spectra $(C_{\ell}^{\gamma_{i}\gamma_{j}},C_{\ell’}^{\gamma_{p}\gamma_{q}})$ in the Gaussian approximation is given by \citep{joachimi:2010}:
\begin{multline}
\mathrm{Cov}[C_{\ell}^{\gamma_{i}\gamma_{j}},C_{\ell’}^{\gamma_{p}\gamma_{q}}]\\
=  \delta_{K}^{\ell\ell’}\frac{2\pi}{(2\ell+1)f_{sky}\Delta\ell} \times \{ C_{\ell}^{\gamma_{i}\gamma_{p}} C_{\ell}^{\gamma_{j}\gamma_{q}}  + C_{\ell}^{\gamma_{i}\gamma_{q}} C_{\ell}^{\gamma_{j}\gamma_{p}} \}  \\
\equiv \mathrm{Cov}^{ijpq}(\ell) \ , \label{eq:13}
\end{multline}
where $\delta_{K}^{\ell\ell’}$ is the Kronecker delta, $f_{sky}$ is the fraction of the sky covered by the survey, and $\Delta\ell$ is the width of multipole bandpowers. We take $f_{sky}$ to be 0.36, corresponding to a 15,000 $\mathrm{deg}^2$ survey such as \textit{Euclid} \citep{euclidprep7:2020}. In a real survey analysis, the sky mask will introduce off-diagonal covariance on scales corresponding to the mask window function. In this case, it is necessary to bin the power spectra and covariance matrix into bandpowers in order for the diagonal covariance assumption to hold. While we do not apply our analysis to the cut sky, we incorporate its effect at the Fisher level by binning our data vector and covariance into bandpowers with a nominal value of $\Delta\ell=10$, and scaling the covariance by $f_{sky}$.
\par
Given our Fisher matrix, fiducial angular power spectra and bias in the power spectrum $\Delta C_{\ell}^{I}$ corresponding to a tomographic bin pair $I \equiv (\gamma_{i},\gamma_{j})$, the resulting contribution to the bias in some cosmological parameter $\theta_{\eta}$ is given by the following \citep{amara:2008,duncan:2020}:
\begin{align}
\Delta\theta_{\eta}^{I} &= \sum_{\tau}(\mathsf{F}^{-1})_{\eta\tau}\sum_{J}\sum_{\ell}(\partial_{\tau}C_{\ell}^{J})(\mathsf{Cov}^{-1})^{J,I}(\ell)\Delta C_{\ell}^{I} \ , \label{eq:14}
\end{align}
where the covariance matrix $\mathrm{Cov}^{J,I}(\ell)$ is given by \eqref{eq:13}. The bias on parameter $\theta_{\eta}$ is then obtained by summing the contributions from each bin:
\begin{equation}
\Delta\theta_{\eta} = \sum_{I}\Delta\theta_{\eta}^{I} \ . \label{eq:15}
\end{equation}
The marginalised $1 \sigma$ uncertainty of parameter $\theta_{\alpha}$ is given by:
\begin{equation}
\sigma_{\alpha} = \sqrt{\left(\mathsf{F}^{-1}\right)_{\alpha \alpha}} \ , \label{eq:16}
\end{equation}
and we calculate the ratio of our calculated bias in parameter $\theta_{\alpha}$ to the corresponding 1-dimensional $1\sigma$ uncertainty as:
\begin{equation}
\mathrm{Sig}[\Delta \theta_{\alpha}] = \frac{\Delta \theta_{\alpha}}{\sigma_{\alpha}} \ . \label{eq:17}
\end{equation}
\par
Henceforth, we refer to the multipole at which we truncate the calculation of the mode-coupling matrix as $\ell_{\mathrm{cut}}$ and the maximum multipole considered in our Fisher analysis as $\ell_{\mathrm{max}}$. In our Fisher analysis, we consider a low $\ell$-cut of $\ell_{\mathrm{min}} = 10$. We apply a range of high $\ell$-cuts in the range $\ell_{\mathrm{max}} = 1500 - 5000$ at the Fisher level, in order to see the effect of including or excluding different scales on the Fisher matrix and parameter biases. Such a cut will be necessary in any real survey, since at high multipoles non-linear modes of the power spectrum contaminate the $C_{\ell}$s, biasing them with respect to the linear theory prediction and degrading the resulting cosmological parameter constraints. According to \cite{euclidprep7:2020}, $\ell_{\mathrm{max}} = 1500$ corresponds to a pessimistic $\ell$-cut, whereas $\ell_{\mathrm{max}} = 5000$ is expected to be optimistic. These cuts are applied both in the computation of the Fisher matrix \eqref{eq:11}, and of the parameter biases, \eqref{eq:14}. We apply our analysis to $\smash{\Delta C_{\ell}^{\gamma\gamma}}$s computed with a range of values for the the peak multipole of the $m$-bias $C_{\ell}$, in the range $\ell_{\mathrm{peak}} = 50 - 5000$, and fix $\sigma_{m}$, the width of the $m$-bias $C_{\ell}$, to $\sigma_{m} = 64$. Additionally, we conduct our analysis keeping $\ell_{\mathrm{max}}$ fixed, and vary $\sigma_{m}$ in the range $64 - 2048$ and $\ell_{\mathrm{peak}}$ in the range $100 - 1100$, in order to test the validity of our choice of $\sigma_{m}$. In our Fisher matrix analysis we consider first the parameter set $\{\Omega_{m},\sigma_{8},w_{0}\}$ (varying $\Omega_{c}$ with $\Omega_{b}$ fixed at the \citet{planck:2018} value) in addition to the extended parameter set $\{\Omega_{m},\sigma_{8},w_{0},w_{a}\}$ and compare the results.
\section{Results}
\label{results}
\subsection{Residual power spectra}
Residual power spectra $\smash{\Delta C_{\ell}^{\gamma_{i}\gamma_{j}}}$ arising due to the spatially-varying $m$-bias are computed by applying the pseudo-$C_{\ell}$ formalism described in \hyperref[pseudo]{section $(2.1)$}. Examples of the fractional residual power spectra $\Delta C_{\ell}^{\gamma_{i}\gamma_{j}}/C_{\ell}^{\gamma_{i}\gamma_{j}}$ are shown in fig. 2, with $\ell_{\mathrm{peak}}$ in the range $50 - 5000$. 
As illustrated in fig. 2, the residual power spectra are generally at least $\sim 4$ orders of magnitude below the fiducial $C_{\ell}$s. As can be seen, the structure of the residual $C_{\ell}$s is highly dependent on the scale of the $m$-bias $C_{\ell}$, $\ell_{\mathrm{peak}}$. There is typically a minimum in the $\Delta C_{\ell}^{\gamma\gamma}/C_{\ell}^{\gamma\gamma}$ at $\ell \lesssim 10$ with an enhancement at higher $\ell$. In the auto-spectra, this enhancement has a gentle gradient and plateau with increasing $\ell$, whereas for the cross-spectra with high $\ell_{\mathrm{peak}}$ the enhancement is more sharply peaked. There is also a smaller drop at still higher $\ell$, with this structure shifting to higher $\ell$ and increasing in strength with increasing $\ell_{\mathrm{peak}}$. The high-$\ell$ smaller minimum of the fractional residual $C_{\ell}$s is typically located at $\ell$ greater than $\ell_{\mathrm{peak}}$ by approximately 500. For lower values of $\ell_{\mathrm{peak}}$ the residual $C_{\ell}$s are enhanced across all $\ell$, and they also vary more smoothly than for higher $\ell_{\mathrm{peak}}$. 
\begin{figure}
\includegraphics[width=0.5\textwidth]{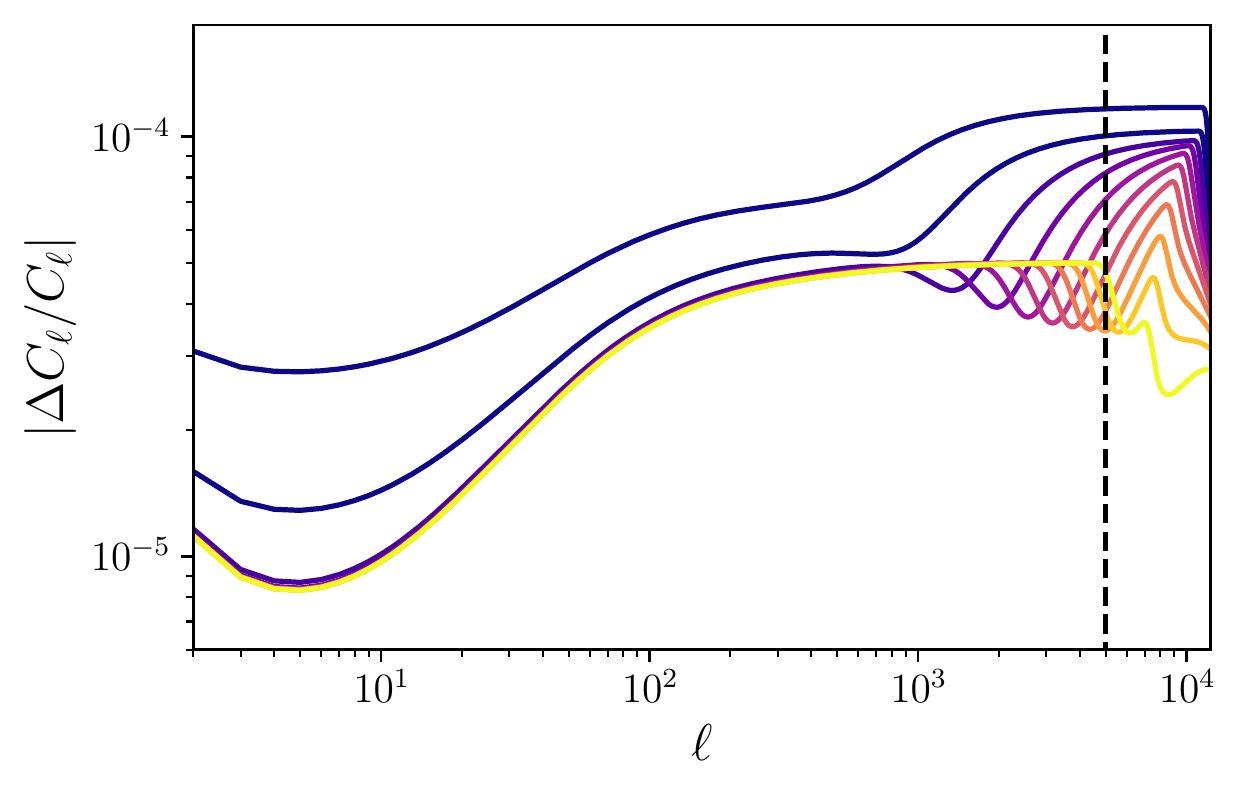}
\includegraphics[width=0.5\textwidth]{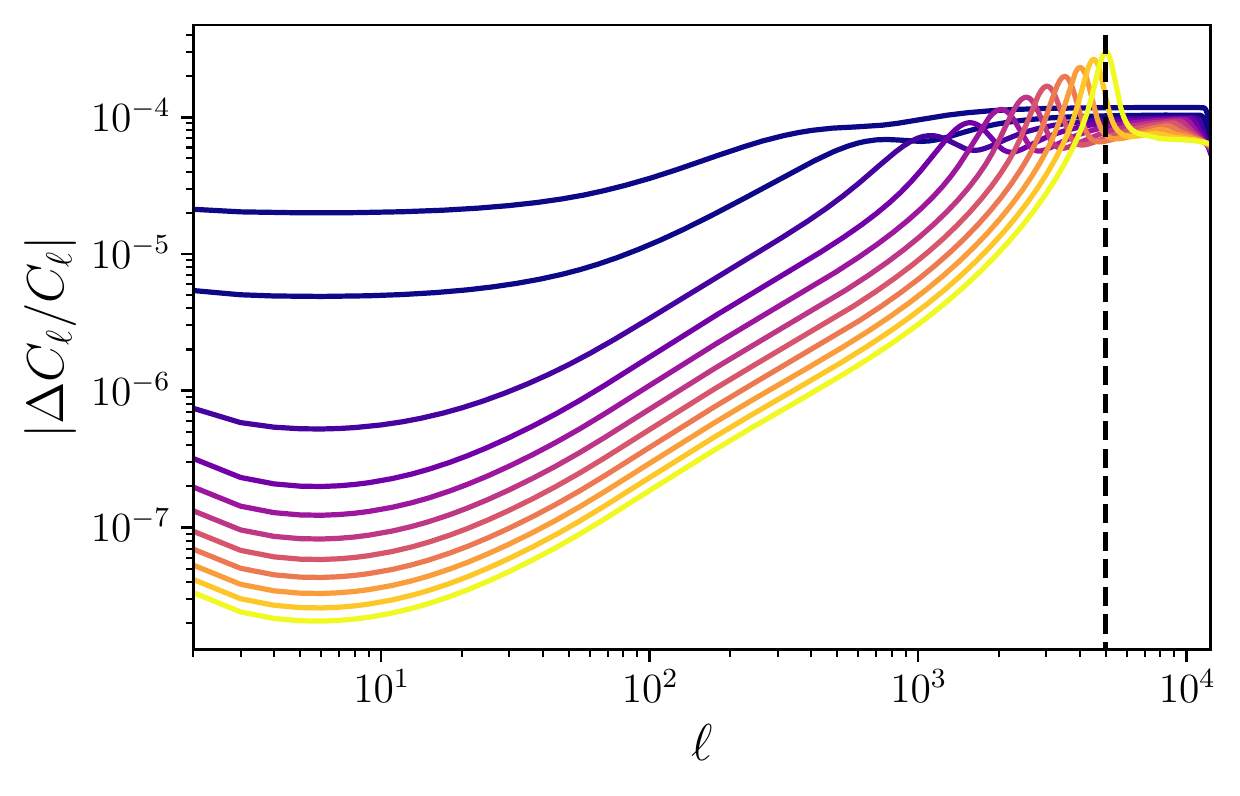}
\includegraphics[width=0.5\textwidth]{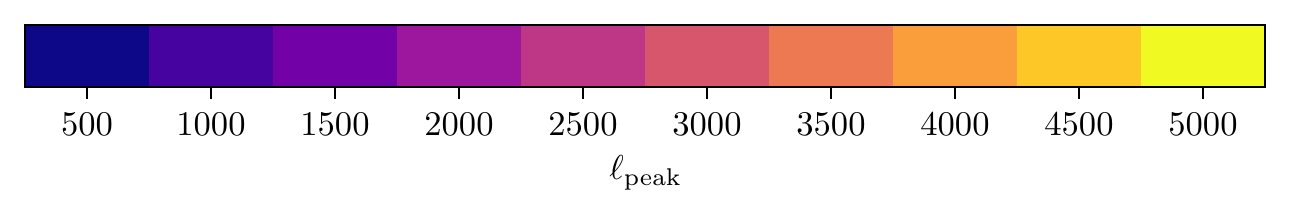}
\caption{Ratio of residual to fiducial angular power spectra as a function of peak multipole of $m$-bias $C_{\ell}$, with $\sigma_{m} = 64$ and $m$-bias rms $= 0.01$. Black dashed line: Maximum value of $\ell_{\mathrm{max}}$ considered in the Fisher analysis. Top panel: autocorrelation of first tomographic redshift bin $\Delta C_{\ell}^{\gamma_{0} \gamma_{0}}/C_{\ell}^{\gamma_{0}\gamma_{0}}$. Bottom panel: cross-correlation of first and fifth tomographic redshift bins $\Delta C_{\ell}^{\gamma_{0} \gamma_{4}}/C_{\ell}^{\gamma_{0}\gamma_{4}}$.}
\end{figure}
\subsubsection{$BB$ power arising due to mode mixing}
We also calculate the $BB$ power spectra arising due to the mode-mixing effect of the spatially-varying $m$-bias, and examples of the $BB$ power spectra are shown in fig. 3, for $\ell_{\mathrm{max}}=5000$ and $\ell_{\mathrm{peak}} = 500$ and $5000$, alongside the corresponding $\Delta C_{\ell}^{EE}$, divided by the fiducial $C_{\ell}^{EE}$. For high $\ell_{\mathrm{peak}}$, the $BB$ power spectra are of the same order of magnitude and follow closely the $\Delta C_{\ell}^{EE}$; for lower $\ell_{\mathrm{peak}}$ the generated $BB$ power spectra follow the $\Delta C_{\ell}^{EE}$ less closely at high and low $\ell$. The observed $BB$ power spectra may therefore be a useful diagnostic of a residual spatially-varying $m$-bias, especially with a small characteristic scale. While this will be complicated by the fact that there are many systematics which may produce measurable shear $B$-modes, such that it will be difficult to isolate the signal arising due to any given systematic in the measured $B$-mode signal, it may be a useful diagnostic if different systematics result in different signature effects in the resulting $BB$ power spectra. As we consider a generic form for the $m$-bias map, we cannot make further conclusions about the viability of the measured $B$-mode signal as a systematics diagnostic in this work.
\begin{figure}
\includegraphics[width=0.5\textwidth]{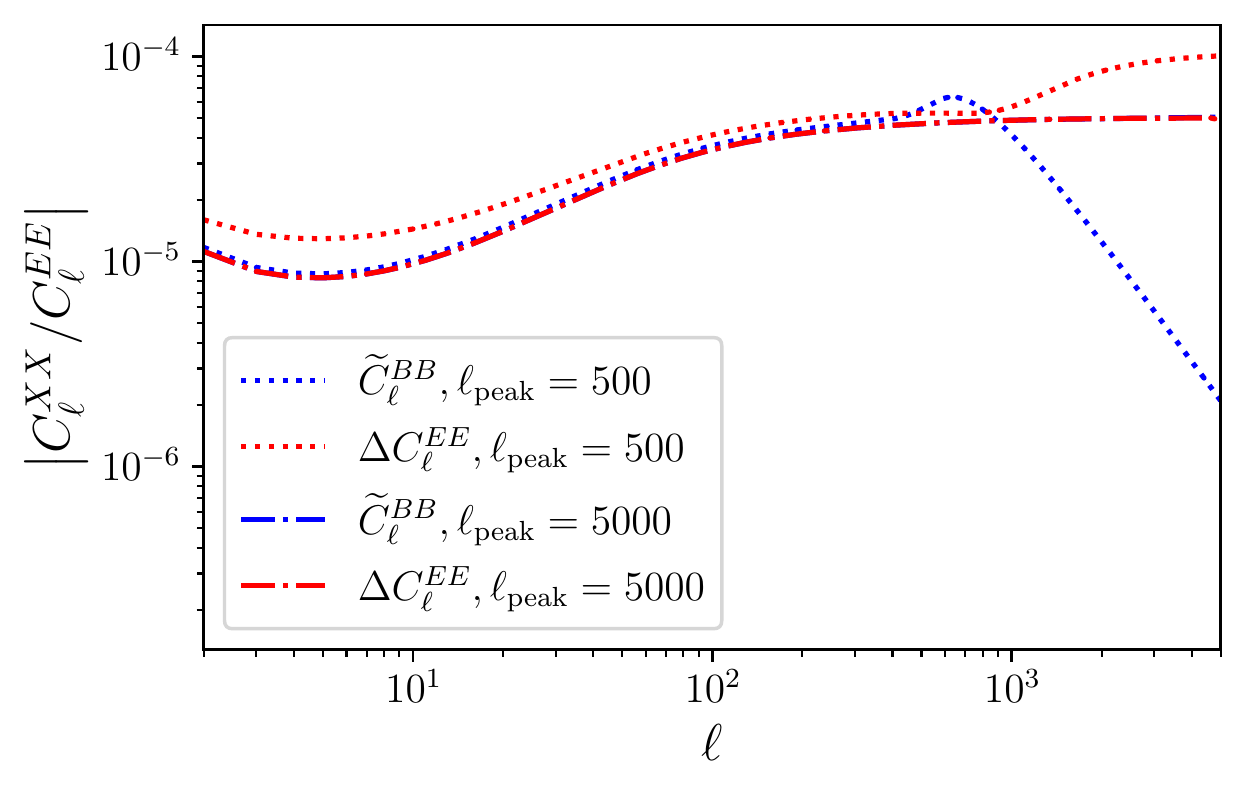}
\includegraphics[width=0.5\textwidth]{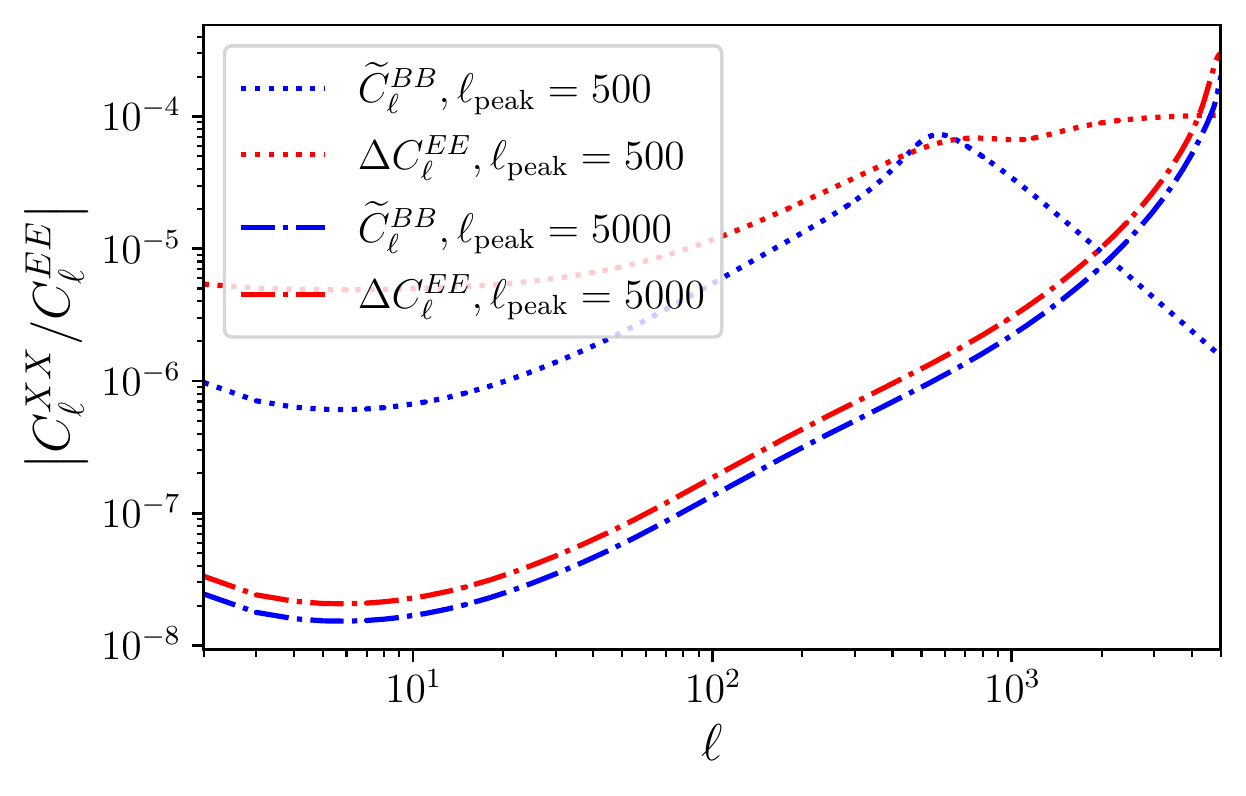}
\caption{Comparison of $BB$ power spectra generated by the mode coupling arising due to the spatially-varying $m$-bias $\widetilde{C}_{\ell}^{BB}$ with the residual $EE$ power spectra $\Delta C_{\ell}^{EE}$, up to $\ell_{\mathrm{max}} = 5000$. Top panel: autocorrelation of first tomographic redshift bin $\Delta C_{\ell}^{\gamma_{0} \gamma_{0}}/C_{\ell}^{\gamma_{0}\gamma_{0}}$. Bottom panel: cross-correlation of first and fifth tomographic redshift bins $\Delta C_{\ell}^{\gamma_{0} \gamma_{4}}/C_{\ell}^{\gamma_{0}\gamma_{4}}$.}
\end{figure}
\subsubsection{Numerical effects in the $\smash{\Delta C_{\ell}^{\gamma\gamma}}$}
As a test of our methods, we vary the multipole at which the calculation of the mode-coupling matrix is truncated, $\ell_{\mathrm{cut}}$. In the top panel of fig. 2, at high $\ell$ and for high $\ell_{\mathrm{peak}}$, the fractional auto-$\Delta C_{\ell}$ curves can be seen to have sharp reductions at $\ell\sim8\times10^{3}-10^{4}$. In order to determine whether this is a numerical effect arising due to the truncation of the calculation of the mode-coupling matrix at $\ell_{\mathrm{cut}}=12288$, the calculation of the $\Delta C_{\ell}$s was also carried out with $\ell_{\mathrm{cut}}=6144$ (corresponding to $N_{\mathrm{side}}=\ell_{\mathrm{cut}}/3=2048$).  We show in fig. 4 the effect of this on examples of the auto- and cross-power spectra for a range of values of $\ell_{\mathrm{peak}}$. As can be seen in the top panel, for the auto-$\Delta C_{\ell}$s, the gentle minimum at intermediate $\ell$ followed by a rise and plateau at higher $\ell$ is robust against changes to $\ell_{\mathrm{cut}}$ and is therefore likely to be a real feature. The same is also true of the sharp peak in the cross-$\Delta C_{\ell}$s. However, for $\ell_{\mathrm{peak}},\ell\gtrsim1.25N_{\mathrm{side}}$, the auto-$\Delta C_{\ell}$s contain a spurious sharp minimum due to the truncation of the calculation of the mode-coupling matrix. Indeed, when $\ell_{\mathrm{peak}}$ exceeds $\sim1.25N_{\mathrm{side}}$ by more than a factor of $\sim 2$, the calculated $\Delta C_{\ell}$s are biased down to smaller $\ell$ (as can be seen in the highest $\ell_{\mathrm{peak}}$ curves in the disparity between the two $\ell_{\mathrm{cut}}$ cases). For $\ell_{\mathrm{cut}}=12288$, $1.25N_{\mathrm{side}}$ corresponds to $\ell\sim5000$ (indicated with a vertical red dashed line in each panel). This justifies the maximum multipole $\ell_{\mathrm{max}}$ at which we truncate our residual power spectra in our Fisher-matrix analysis; in addition to corresponding to the optimistic \textit{Euclid} setting \citep{euclidprep7:2020}, this also prevents contamination of our analysis by edge effects. Due to this effect, we also neglect $\ell_{\mathrm{peak}}>5000$ in our Fisher-matrix analysis; this corresponds to scales of $\lesssim2$ arcminutes, so that we still probe scales on which spatially-varying $m$-bias might be expected to arise due to PSF errors. This effect was also tested for the other tomographic power spectra with similar results.
\begin{figure}
\includegraphics[width=0.5\textwidth]{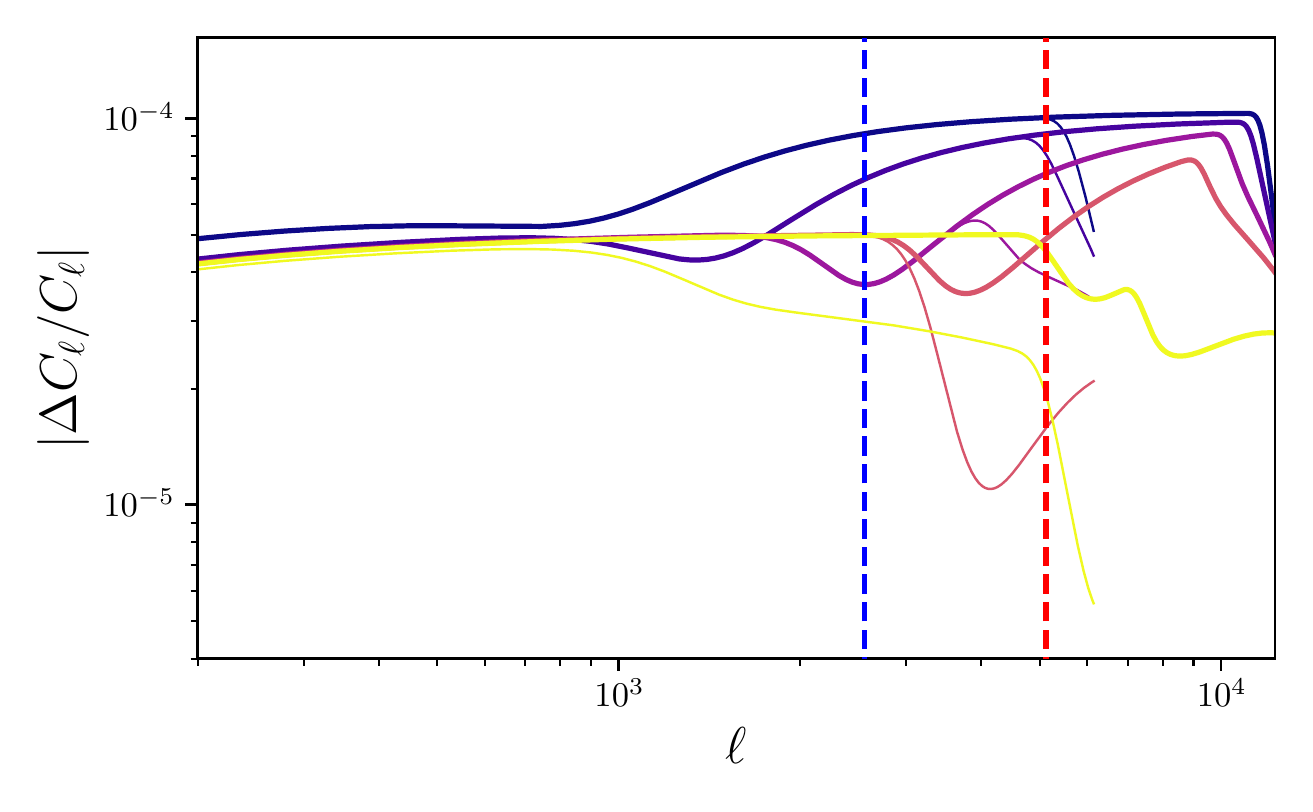}
\includegraphics[width=0.5\textwidth]{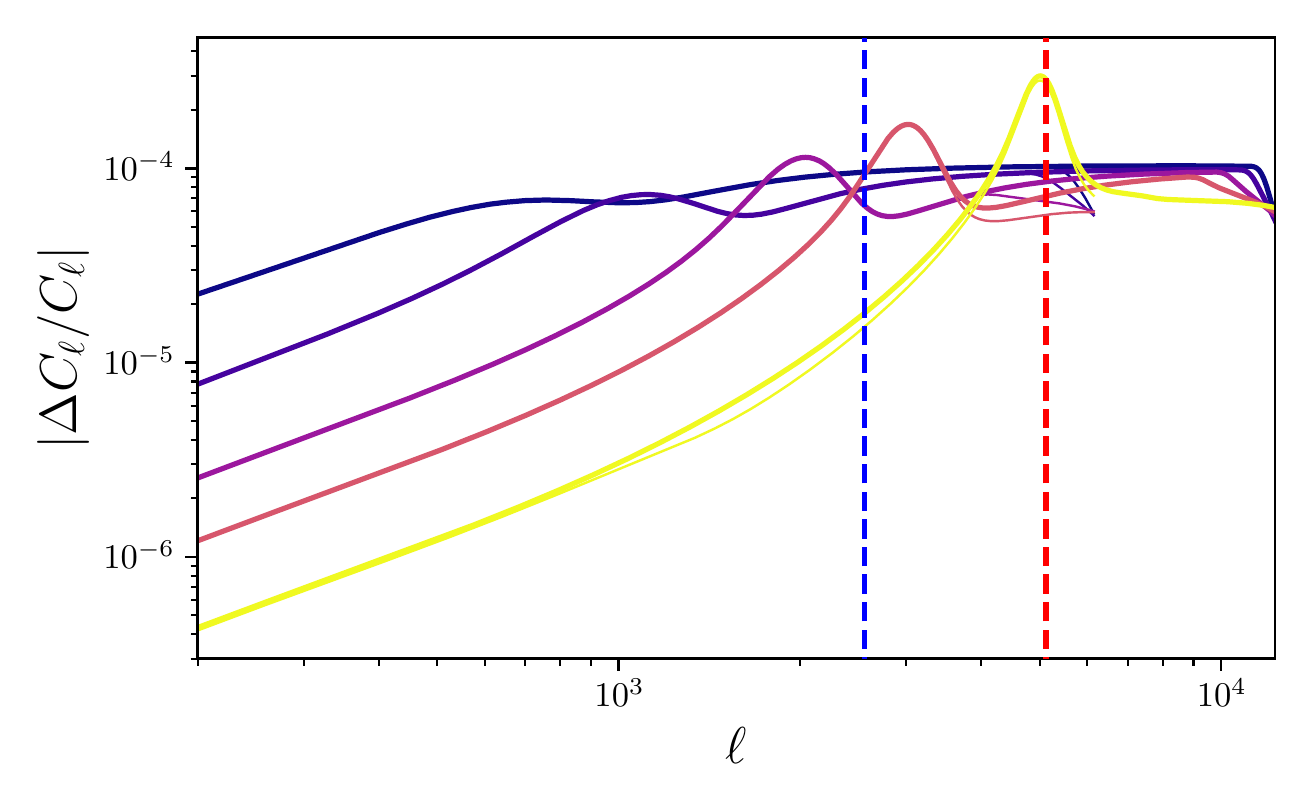}
\includegraphics[width=0.5\textwidth]{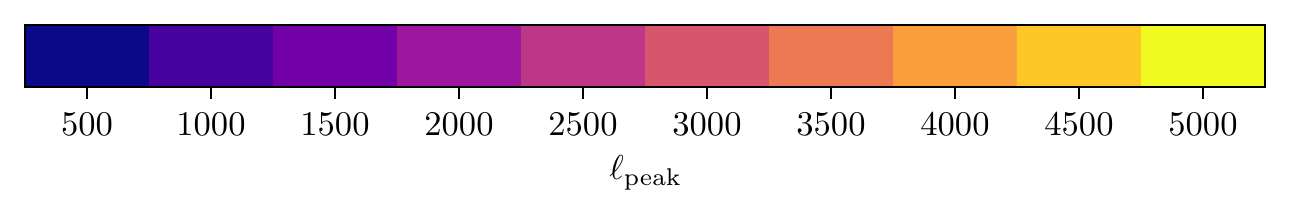}
\caption{Ratio of residual to fiducial angular power spectra as a function of peak multipole of $m$-bias $C_{\ell}$, with $\sigma_{m} = 64$ and $m$-bias rms $= 0.01$, showing the effect of the truncation of the calculation of the mode-coupling matrix at a particular multipole $\ell_{\mathrm{cut}}$. Bold lines: $\ell_{\mathrm{cut}} = 3N_{\mathrm{side}} = 12288$. Thin lines: $\ell_{\mathrm{cut}} = 3N_{\mathrm{side}} = 6144$. Vertical dashed lines correspond to the maximum threshhold values of $\ell_{\mathrm{max}}$ that could be considered in a subsequent Fisher analysis, equal to $1.25N_{\mathrm{side}}$; blue for the $\ell_{\mathrm{cut}} = 6144$ case and red for the $\ell_{\mathrm{cut}} = 12288$ case which we consider. Top panel: autocorrelation of first tomographic redshift bin $\Delta C_{\ell}^{\gamma_{0} \gamma_{0}}/C_{\ell}^{\gamma_{0}\gamma_{0}}$. Bottom panel: cross-correlation of first and fifth tomographic redshift bins $\Delta C_{\ell}^{\gamma_{0} \gamma_{4}}/C_{\ell}^{\gamma_{0}\gamma_{4}}$.}
\end{figure}
\subsection{Bias in cosmological parameters}
Using the Fisher matrix analysis described in \hyperref[fisher]{section $(2.3)$}, biases due to shifts in the $\smash{C_{\ell}^{\gamma_{i}\gamma_{j}}}$ are forecast for the base cosmological parameter set $\{ \Omega_{m}, \sigma_{8}, w_{0} \}$ (varying $\Omega_{c}$ in our Fisher-matrix analysis, with $\Omega_{b}$ fixed at the \citet{planck:2018} value). We apply a high-$\ell$ cut $\ell_{\mathrm{max}}$ in both the computation of the Fisher matrix \eqref{eq:11} and of the parameter biases \eqref{eq:14}, and repeat the analysis for a range of values of $\ell_{\mathrm{max}}$. In fig. 5 we show the ratio of our calculated parameter biases to the $1\sigma$ uncertainty as a function of $\ell_{\mathrm{peak}}$ and $\ell_{\mathrm{max}}$ with $\sigma_{m}$ fixed at 64. We next consider the extended parameter set $\{\Omega_{m},\sigma_{8},w_{0},w_{a}\}$, and in fig. 6 show the ratio of our parameter biases to the $1\sigma$ uncertainty as a function of $\ell_{\mathrm{peak}}$ and $\ell_{\mathrm{max}}$ with $\sigma_{m}$, the width of the $m$-bias $C_{\ell}$, fixed at 64. We also conducted our analysis keeping $\ell_{\mathrm{max}}$ fixed, and varying $\sigma_{m}$ in the range $64 - 2048$, and $\ell_{\mathrm{peak}}$ in the range $100 - 1100$. We found that for all parameters considered, the biases and their ratios to the $1\sigma$ uncertainty were insensitive to the value of $\sigma_{m}$ for $\sigma_{m} \lesssim 500$, justifying our choice of $\sigma_{m} = 64$ for our main analysis. Larger values of $\sigma_{m}$ for a given value of $m$-bias rms resulted in reduced cosmological parameter bias.
\par
In Table 2, we report the maximum absolute values of the bias-uncertainty ratios obtained in parameter biases, for the base parameter set of $\{ \Omega_{m}, \sigma_{8}, w_{0} \}$. We report bias-uncertainty ratios for three values of $\ell_{\mathrm{max}}$, corresponding to pessimistic and optimistic settings for a  \textit{Euclid}-style survey \citep{euclidprep7:2020} (note that for the 3-parameter case, the highest ratio of bias to 1$\sigma$ uncertainty in $\sigma_{8}$ of 0.032 is achieved for $\ell_{\mathrm{max}} = 3000$). In table 3, we report the same, for the extended parameter set of $\{ \Omega_{m}, \sigma_{8}, w_{0}, w_{a} \}$.
\par
For both parameter sets, there are two clear trends; firstly, that the parameter bias-uncertainty ratio is greater when higher multipoles $\ell$ are included in the Fisher analysis; and secondly, that the $m$-bias generally has a greater influence when it is peaked at smaller $\ell$. Furthermore, the inclusion of the parameter $w_{a}$ in the Fisher matrix has resulted in more significant biases for all parameters, due to the additional parameter degeneracy in the $C_{\ell}$s arising when more parameters are included in the Fisher analysis. There is additionally a trend of subsidiary peaks in the bias-uncertainty ratios at higher $\ell_{\mathrm{peak}}$ for all parameters, which is typically strongest where $\ell_{\mathrm{peak}}$ is less than $\ell_{\mathrm{max}}$ by $\sim$ 500, though these peaks become suppressed for the highest values of $\ell_{\mathrm{max}}$. These occur because a given value of $\ell_{\mathrm{max}}$ will roughly coincide with the high-$\ell$ minimum of the $\smash{\Delta C_{\ell}^{\gamma\gamma}}$ for which $\ell_{\mathrm{peak}} \sim \ell_{\mathrm{max}} - 500$. Therefore, for smaller values of $\ell_{\mathrm{peak}}$ than this, for which the structure in the $\smash{\Delta C_{\ell}^{\gamma\gamma}}$ is shifted to lower $\ell$, less of the $\smash{\Delta C_{\ell}^{\gamma\gamma}}$ that is included in the bias calculation will consist of the low-$\ell$ plateau, and more will be in the higher-$\ell$ trough; whereas for larger values of $\ell_{\mathrm{peak}}$, the $\smash{\Delta C_{\ell}^{\gamma\gamma}}$ is generally suppressed. In both cases, the resulting impact in the inferred cosmology is smaller, so we get a peak in parameter bias at this value of $\ell_{\mathrm{peak}}$. At the lowest values of $\ell_{\mathrm{peak}}$, the $\smash{\Delta C_{\ell}^{\gamma\gamma}}$ is generally enhanced, resulting in the low-$\ell_{\mathrm{peak}}$ rise in parameter bias and bias-uncertainty ratio.
\begin{table}
\centering
\caption{Maximum absolute bias$/1\sigma$ for pessimistic (1500) and optimistic (5000) values of $\ell_{\mathrm{max}}$, base parameter set (2 s.f.).}
\begin{tabular}{| c || c c |}
\hline\hline
\multicolumn{1}{|c|}{} & \multicolumn{2}{|c|}{Bias$/1\sigma$}  \\
\hline
Parameter & $\ell_{\mathrm{max}} = 1500$ & $\ell_{\mathrm{max}} = 5000$  \\
\hline\hline
$\Omega_{m}$ &  $0.0059$ & $0.052$ \\
$\sigma_{8}$ & $0.020$ & $0.017$ \\
$w_{0}$ & $0.013$ & $0.068$ \\
\hline\hline
\end{tabular}
\end{table}
\begin{table}
\centering
\caption{Maximum absolute bias$/1\sigma$ for pessimistic (1500) and optimistic (5000) values of $\ell_{\mathrm{max}}$, extended parameter set (2 s.f.).}
\begin{tabular}{| c || c c |}
\hline\hline
\multicolumn{1}{|c|}{} & \multicolumn{2}{|c|}{Bias$/1\sigma$}  \\
\hline
Parameter & $\ell_{\mathrm{max}} = 1500$ & $\ell_{\mathrm{max}} = 5000$  \\
\hline\hline
$\Omega_{m}$ &  $0.014$ & $0.10$ \\
$\sigma_{8}$ & $0.0057$ & $0.063$ \\
$w_{0}$ & $0.025$ & $0.13$ \\
$w_{a}$ & $0.021$ & $0.11$ \\
\hline\hline
\end{tabular}
\end{table}
\begin{figure}
\includegraphics[width=0.5\textwidth]{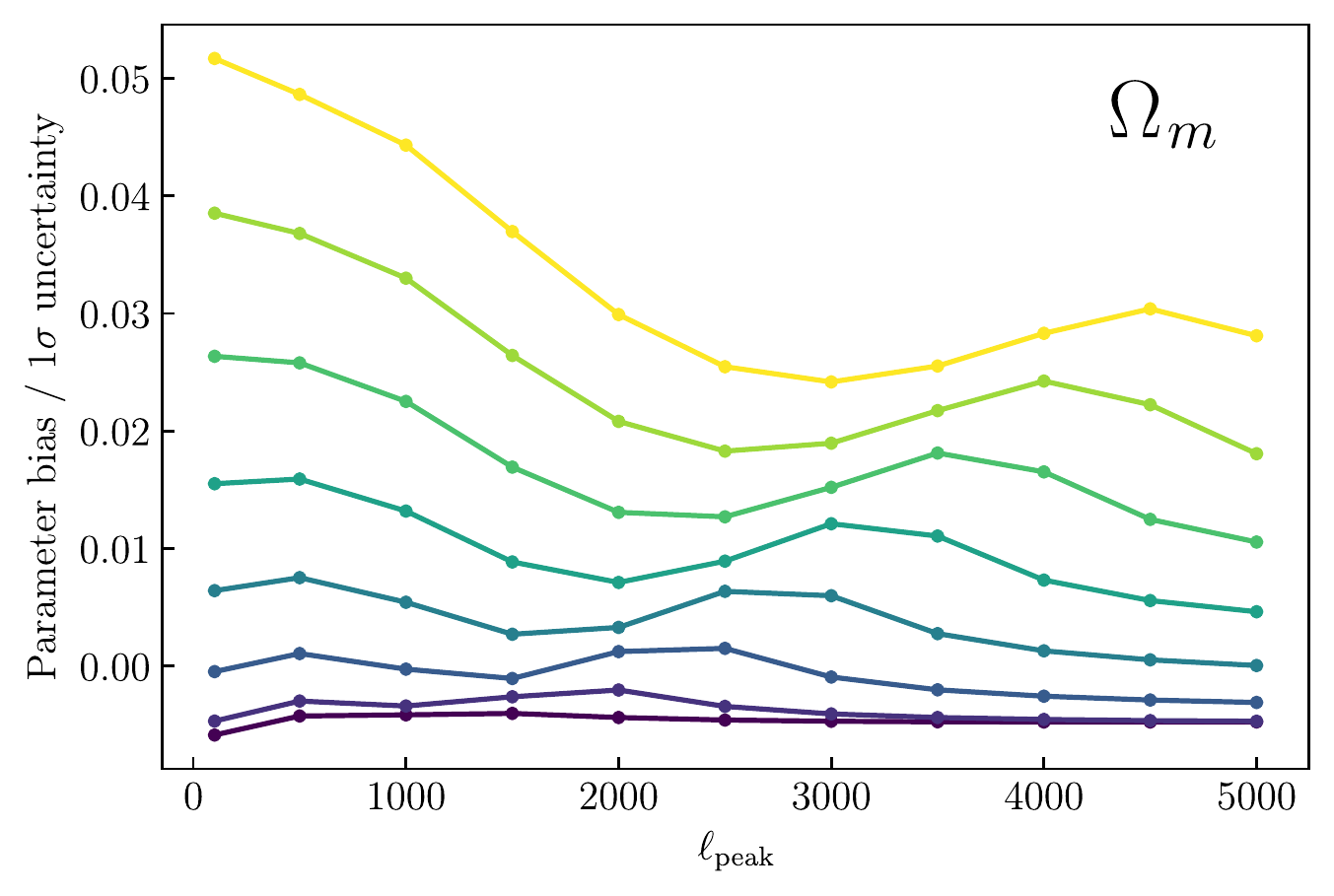}
\includegraphics[width=0.5\textwidth]{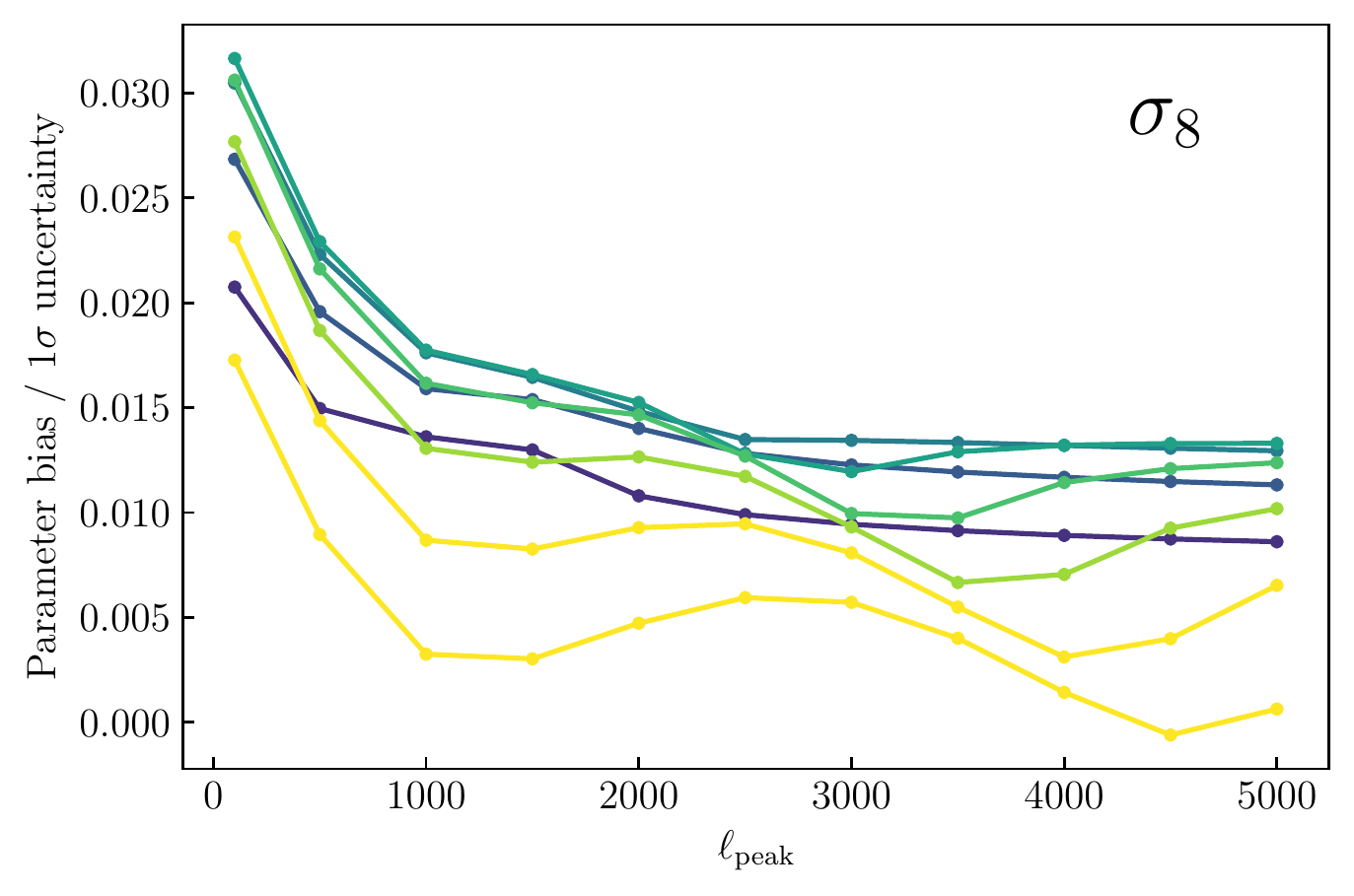}
\includegraphics[width=0.5\textwidth]{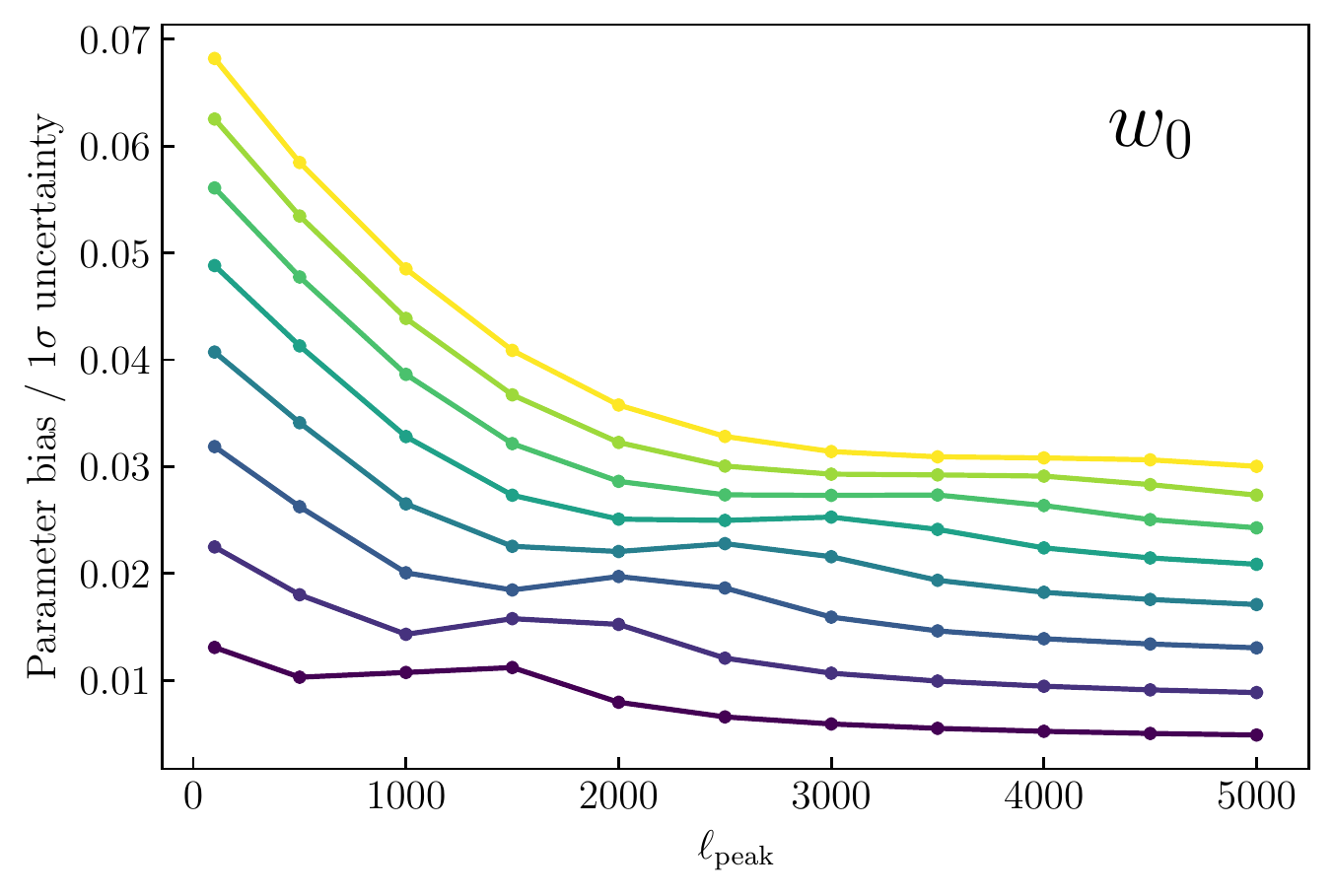}
\includegraphics[width=0.5\textwidth]{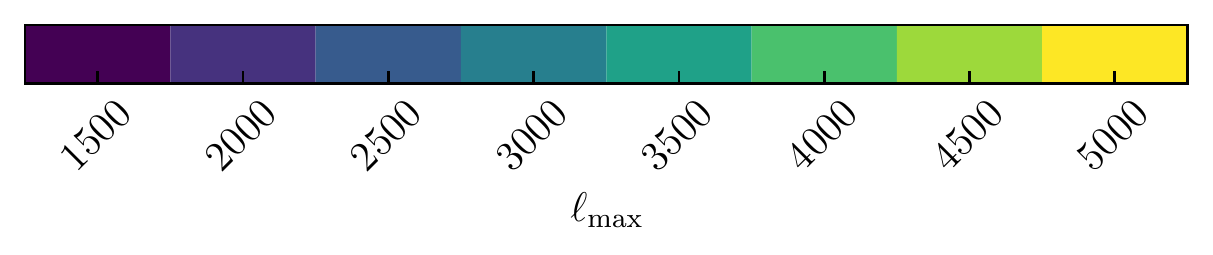}
\caption{Parameter bias-uncertainty ratio in the case that $\{\Omega_{m},\sigma_{8},w_{0}\}$ are included in the Fisher matrix. \textbf{Upper panel:} Bias-uncertainty ratio for $\Omega_{m}$ as a function of $\ell_{\mathrm{peak}}$ and $\ell_{\mathrm{max}}$.
\textbf{Middle panel:} Bias-uncertainty ratio for $\sigma_{8}$ as a function of $\ell_{\mathrm{peak}}$ and $\ell_{\mathrm{max}}$.
\textbf{Lower panel:} Bias-uncertainty ratio for $w_{0}$ as a function of $\ell_{\mathrm{peak}}$ and $\ell_{\mathrm{max}}$.}
\end{figure}
\begin{figure*}
\begin{tabular}{cc}
\includegraphics[width=0.5\textwidth]{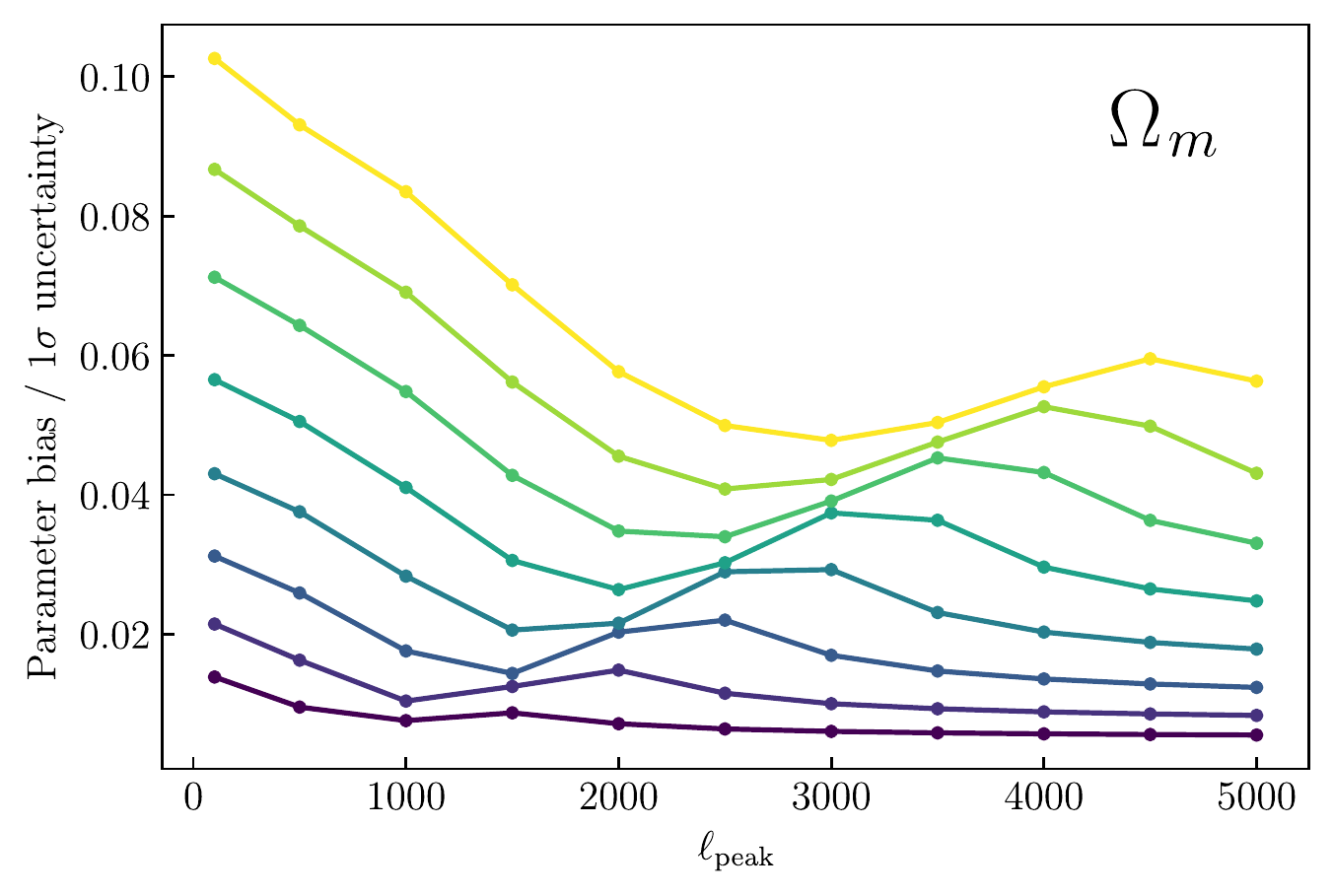} & \includegraphics[width=0.5\textwidth]{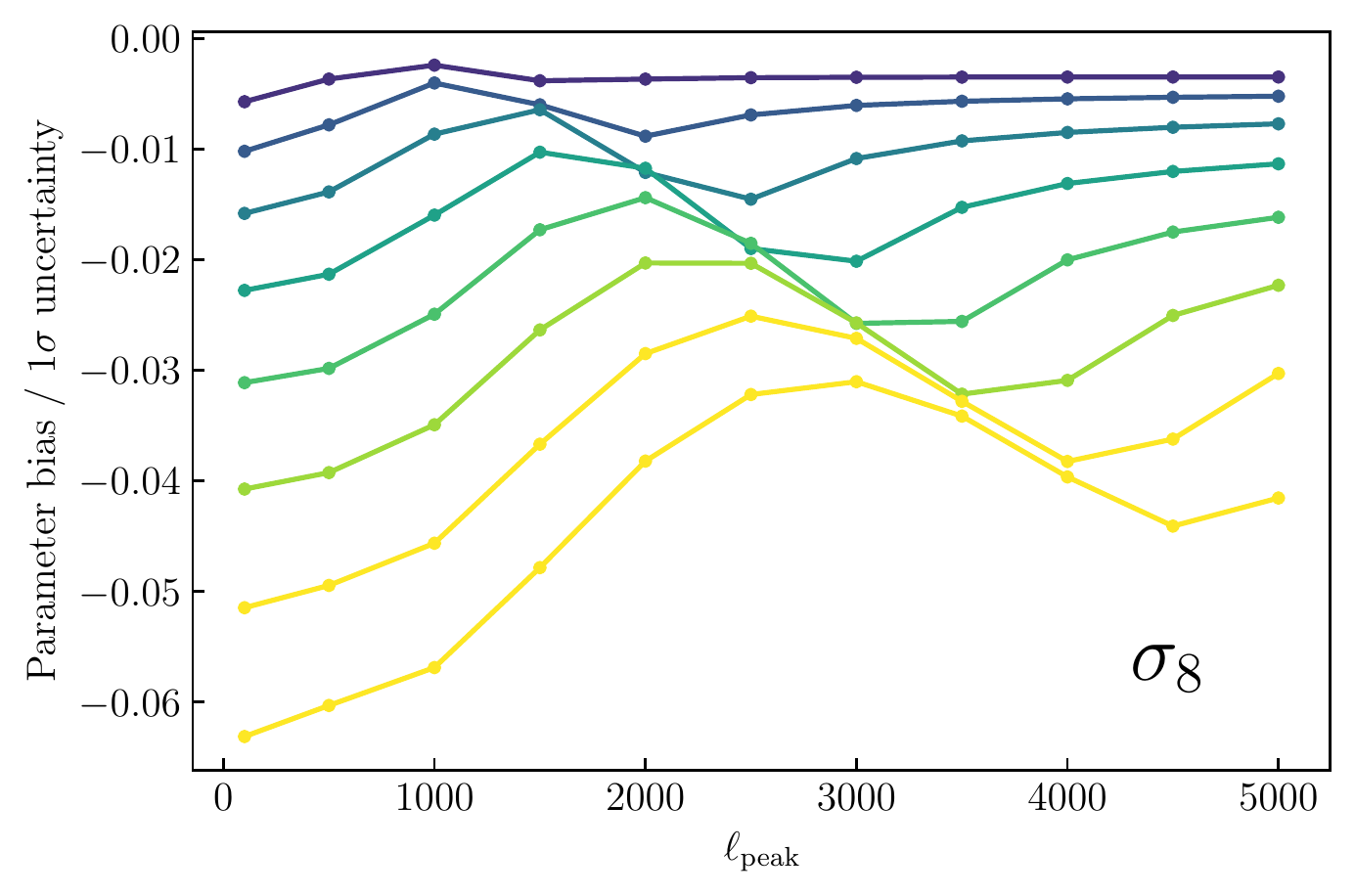} \\
\includegraphics[width=0.5\textwidth]{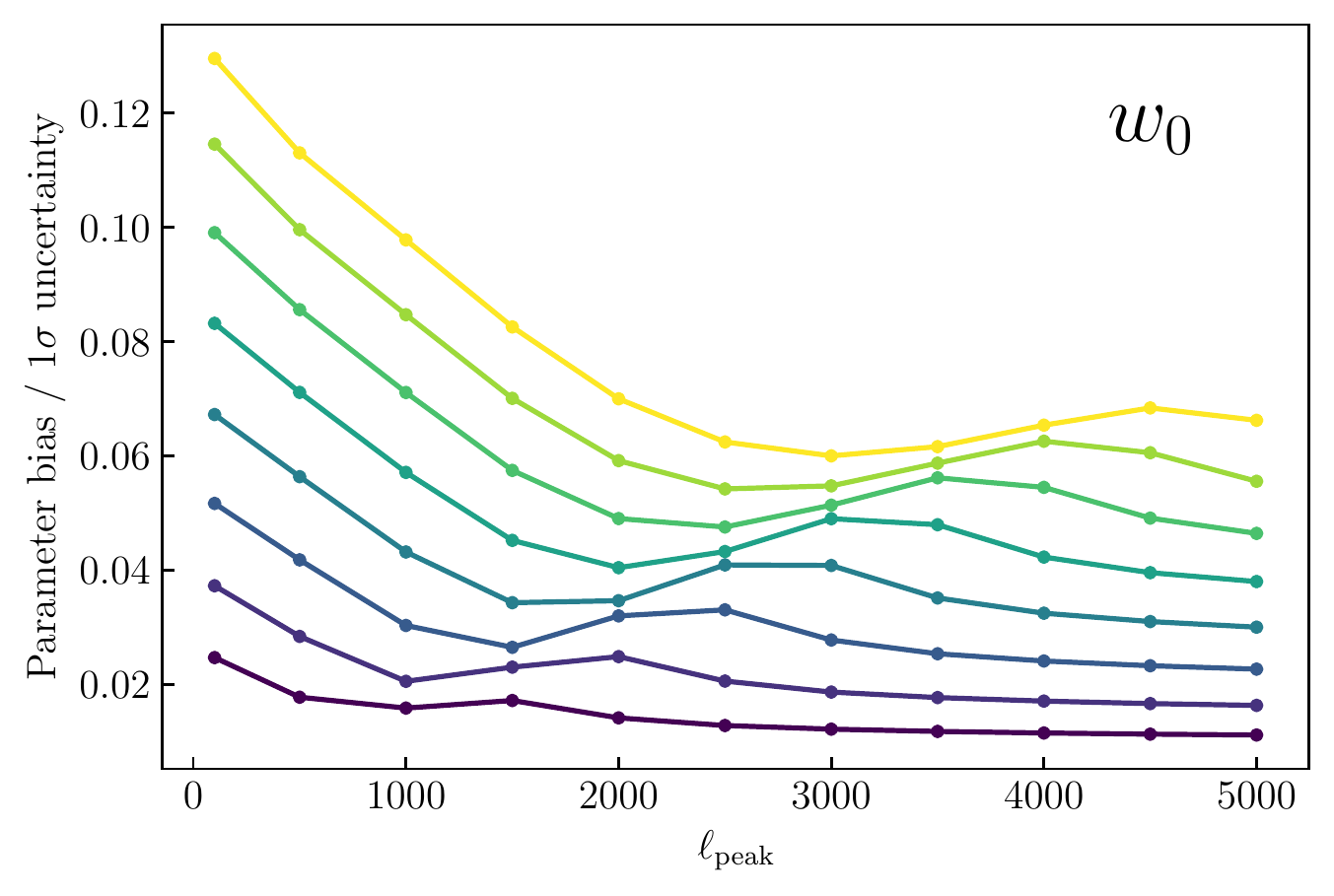} &
\includegraphics[width=0.5\textwidth]{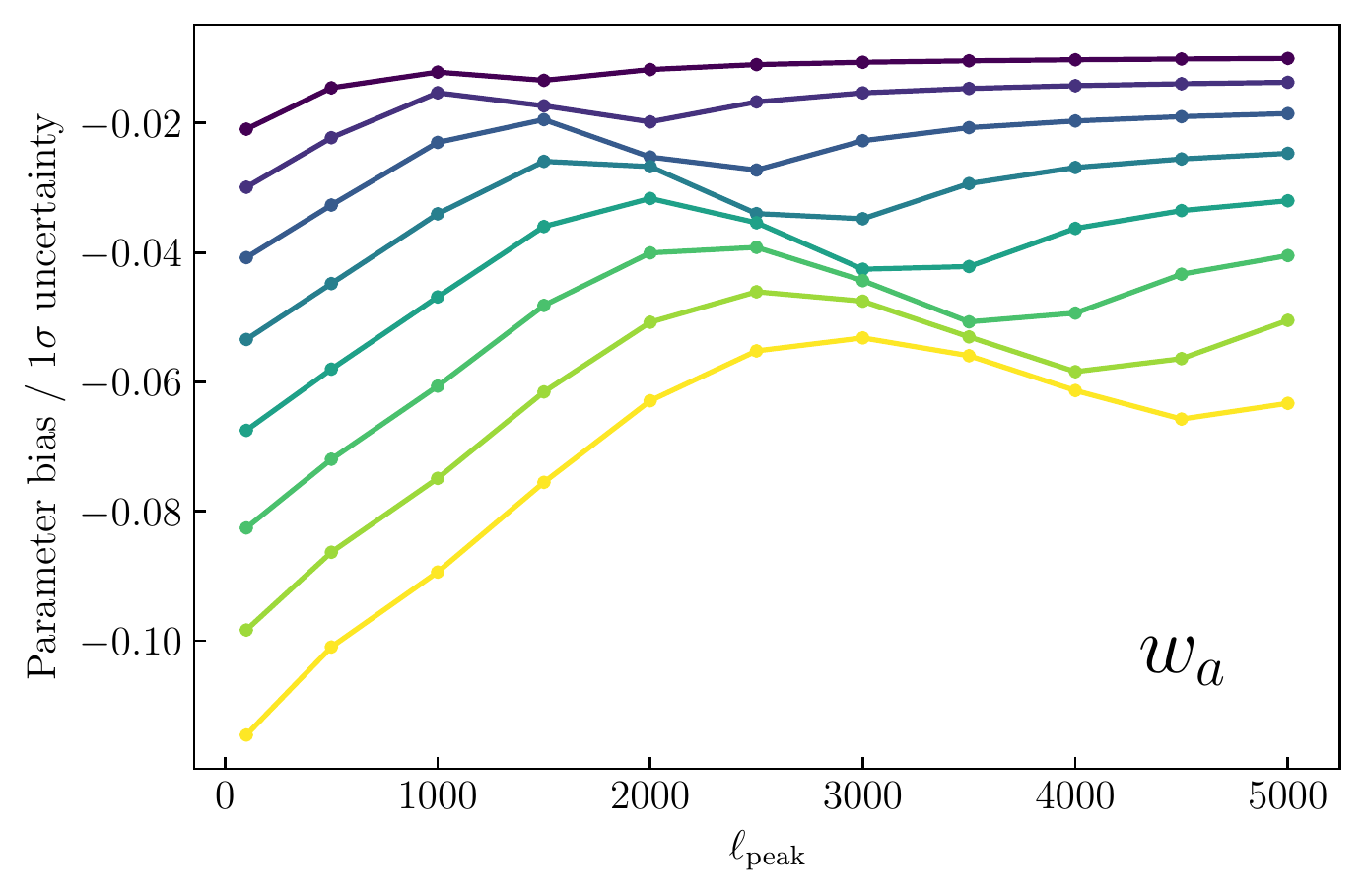}
\end{tabular}
\includegraphics[width=0.5\textwidth]{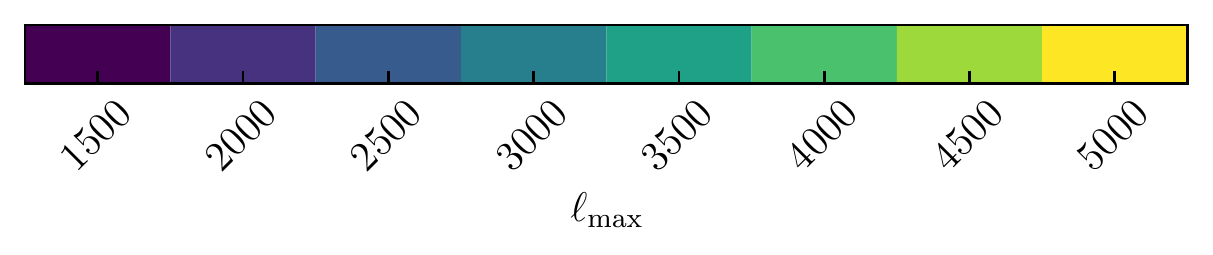}
\caption{Parameter bias-uncertainty ratio in the case that $\{\Omega_{m},\sigma_{8},w_{0},w_{a}\}$ are included in the Fisher matrix. \textbf{Upper-left panel:} Bias-uncertainty ratio for $\Omega_{m}$ as a function of $\ell_{\mathrm{peak}}$ and $\ell_{\mathrm{max}}$.
\textbf{Upper-right panel:} Bias-uncertainty ratio for $\sigma_{8}$ as a function of $\ell_{\mathrm{peak}}$ and $\ell_{\mathrm{max}}$.
\textbf{Lower-left panel:} Bias-uncertainty ratio for $w_{0}$ as a function of $\ell_{\mathrm{peak}}$ and $\ell_{\mathrm{max}}$.
\textbf{Lower-right panel:} Bias-uncertainty ratio for $w_{a}$ as a function of $\ell_{\mathrm{peak}}$ and $\ell_{\mathrm{max}}$.}
\end{figure*}
\subsection{Dependence of parameter bias on $m$-bias rms}
We also investigate how our forecast parameter biases vary with the amplitude of the variations of the $m$-bias map. In fig. 7 we show the $m$-bias rms which results in parameter bias-uncertainty ratios of $\Delta\theta/\sigma = 0.1$, $0.3$ and $1$ as a function of $\ell_{\mathrm{peak}}$, for the base parameter set of $\{ \Omega_{m}, \sigma_{8}, w_{0} \}$, with $\ell_{\mathrm{max}} = 5000$. In fig. 8 we show this for the extended parameter set of $\{ \Omega_{m}, \sigma_{8}, w_{0}, w_{a} \}$. As can be seen, a given bias-uncertainty ratio results from a smaller rms for smaller values of $\ell_{\mathrm{peak}}$, and has a weakly increasing dependence on the number of parameters included in the Fisher matrix. For both parameter sets, the bias-uncertainty ratio is $\sim0.1$ when the rms$=0.01$, $\sim0.3$ when the rms$=0.02-0.03$, and the parameter biases reach the statistical errors when the rms$=0.03-0.06$, depending on $\ell_{\mathrm{peak}}$. 
\begin{figure}
\includegraphics[width=0.5\textwidth]{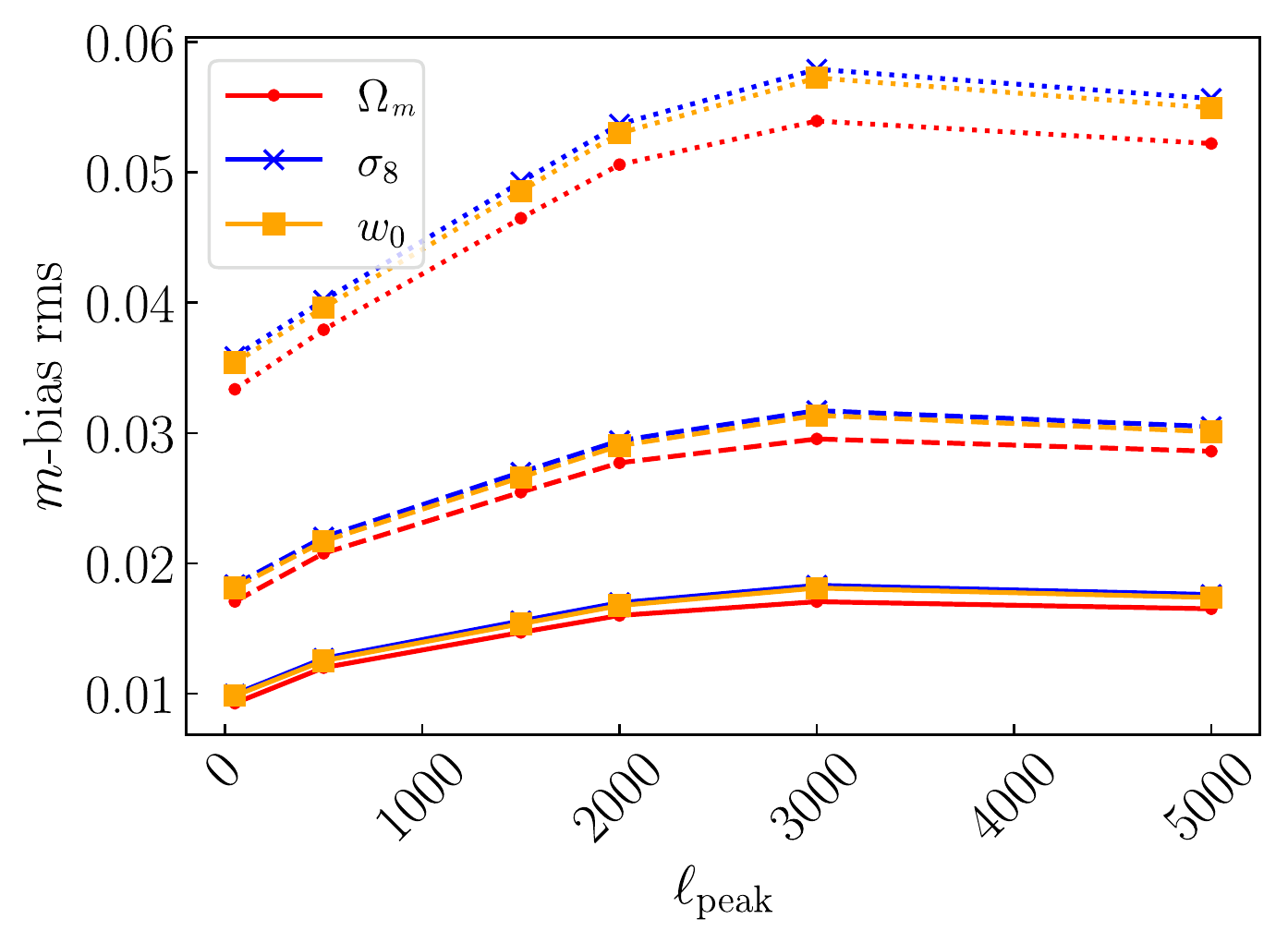}
\caption{$m$-bias rms which results in selected values of parameter bias-uncertainty ratio for the base parameter set $\{\Omega_{c}, \sigma_{8}, w_{0}\}$, as a function of $\ell_{\mathrm{peak}}$. Solid line: $m$-bias rms which results in bias-uncertainty ratio of $0.1$. Dashed line: $m$-bias rms which results in bias-uncertainty ratio of $0.3$. Dotted line: $m$-bias rms which results in bias-uncertainty ratio of $1$.}
\end{figure}
\begin{figure}
\includegraphics[width=0.5\textwidth]{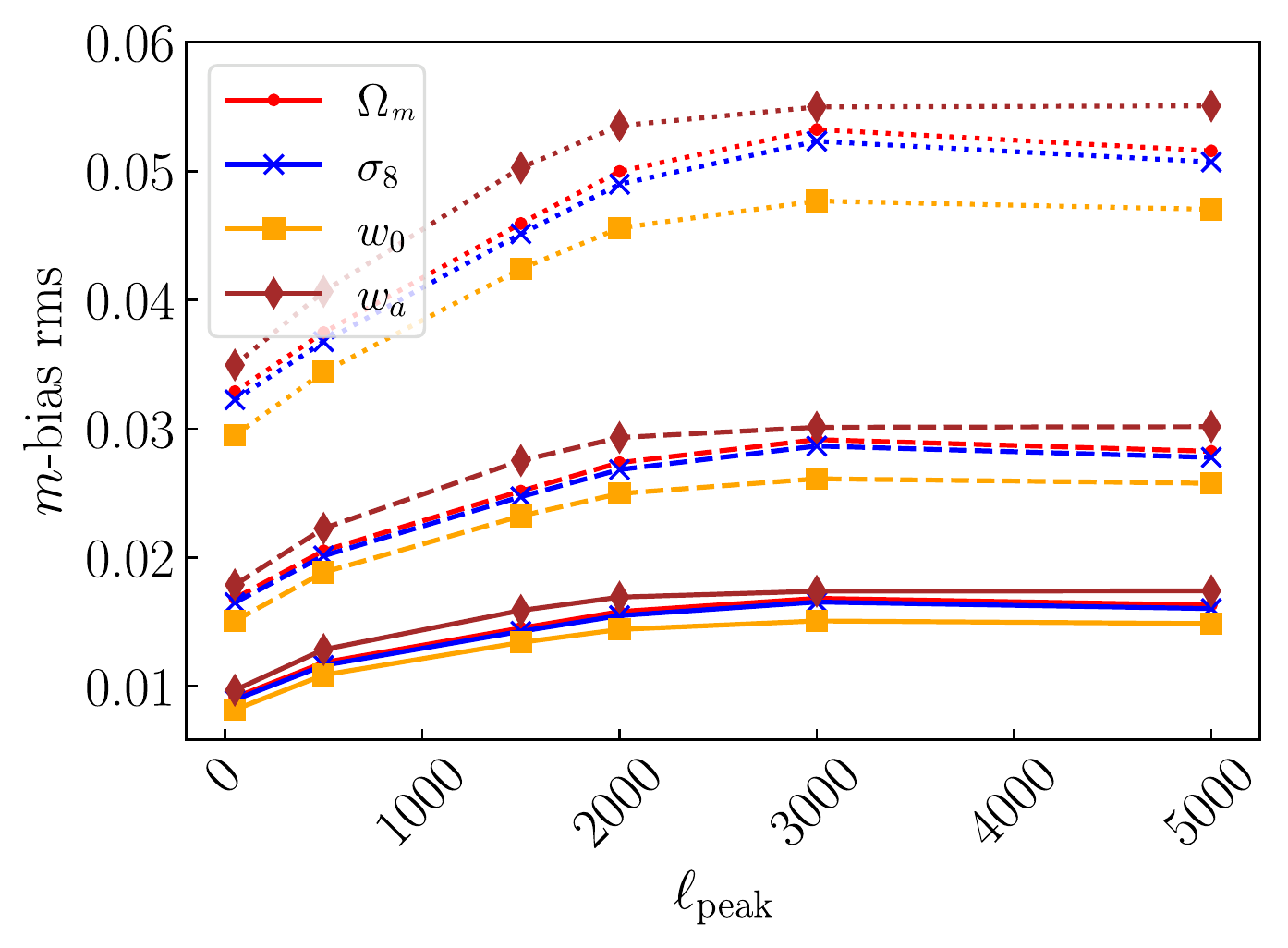} 
\caption{$m$-bias rms which results in selected values of parameter bias-uncertainty ratio for the extended parameter set $\{\Omega_{c}, \sigma_{8}, w_{0}, w_{a}\}$, as a function of $\ell_{\mathrm{peak}}$. Solid line: $m$-bias rms which results in bias-uncertainty ratio of $0.1$. Dashed line: $m$-bias rms which results in bias-uncertainty ratio of $0.3$. Dotted line: $m$-bias rms which results in bias-uncertainty ratio of $1$.}
\end{figure}
\section{Discussion}
\label{discussion}
\subsection{Residual power spectra}
As can be seen in the fractional $\smash{\Delta C_{\ell}^{\gamma\gamma}}$ shown in fig. 2, an $m$-bias with some particular $\ell_{\mathrm{peak}}$ can introduce a bias in the $C_{\ell}$s with generally consistent magnitude on a range of scales, due to mode-mode coupling; therefore, any analysis of the effect of an $m$-bias on the $C_{\ell}$s must not be limited only to the scales on which the $m$-bias varies. Note that as the effect of the spatially-varying $m$-bias on the power spectrum is more complicated than a constant scaling, the impact on the cosmological parameters due to marginalisation over a constant bias parameter will not be an effective means of mitigation. In a real survey, the $C_{\ell}$ of the $m$-bias map will not be not known a priori, and will probably not be easily describable with a small number of parameters as in the case studied here. As a result, it will be difficult to marginalise over the effect of a spatially-varying $m$-bias in cosmological parameter estimation, because the number of required nuisance parameters will be too large. Due to this we have not considered the effect of $m$-bias marginalisation on cosmological parameter biases or figures of merit in the present work.
\par
The fractional residual shear angular power spectra we find reach a maximum of $\sim10^{-4}$. \cite{kitching:2019,kitching:2020} also calculated residuals between the full analytic shear $C_{\ell}$s, which included spatially-varying shear bias, and an analytic linear approximation, which only included mean and spatially-constant terms, up to a maximum multipole of 64. \cite{kitching:2020} also calculated residuals between forward-modelled shear $C_{\ell}$s, which included spatially-varying shear bias, and the linear approximation, up to a maximum multipole of 2048, and found differences which were also $\sim$4 orders of magnitude smaller than the fiducial power spectra. We therefore find similar results at the order-of-magnitude level in the regions of $\ell$-space in which our results overlap.
\subsection{Bias in cosmological parameters}
As shown in figs. 5 and 6, the parameter bias-uncertainty ratio generally increases with decreasing $\ell_{\mathrm{peak}}$ and increasing $\ell_{\mathrm{max}}$. The increase in bias-uncertainty ratio (and indeed in bias) with $\ell_{\mathrm{max}}$ occurs because the $C_{\ell}$s are biased on all scales and so including contributions from additional scales will increase the size of the parameter bias. Indeed, while including higher multipoles in the analysis results in a smaller parameter uncertainty, this effect is small compared with the increase in parameter bias with $\ell_{\mathrm{max}}$. The greater influence of an $m$-bias with small $\ell_{\mathrm{peak}}$ arises due to the generally larger $\smash{\Delta C_{\ell}^{\gamma\gamma}}$ for smaller $\ell_{\mathrm{peak}}$. $m$-bias maps with different $\ell_{\mathrm{peak}}$ also affect each parameter differently; each parameter exhibits different variation in the parameter bias with $\ell_{\mathrm{peak}}$ and $\ell_{\mathrm{max}}$, due to sensitivity of different parts of the shear power spectra to each parameter. The scale-dependence of the parameter biases which result from the spatial variations of the $m$-bias field, which come in through the $\ell_{\mathrm{peak}}$-dependence, demonstrates that the bias in shear power spectra and in cosmological parameters varies with the shape of the power spectrum of $m$-bias field. This shows that the impact of spatial variations of the $m$-bias on cosmological parameter estimation can not be fully mitigated by marginalisation, unless a parametric form of power spectrum of the $m$-bias field is known, which will be unlikely in practice.
\par
Maximum values of parameter bias-uncertainty ratio obtained for the pessimistic and optimistic settings for $\ell_{\mathrm{max}}$ \citep{euclidprep7:2020} are shown in tables $2$ and $3$, for the base and extended parameter sets respectively. Larger parameter bias-uncertainty ratios are obtained when more free parameters are included in the Fisher analysis. In the case of the optimistic cutoff multipole for a \textit{Euclid}-style survey, corresponding to $\ell_{\mathrm{max}} = 5000$, the maximum bias-uncertainty ratios are at about $0.05$ and $0.1$ for the base and extended parameters sets, respectively. In the case of $w_{0}$, for which the maximum bias-uncertainty ratios for $\ell_{\mathrm{max}} = 5000$ is $0.12$, this may still represent a non-negligible contribution to the total systematic error budget; this is potentially relevant in light of the fact that one of the main science goals of \textit{Euclid} and other stage-IV surveys is the investigation of dark energy and the accelerated expansion \citep{euclid:2011}. In the pessimistic setting of $\ell_{\mathrm{max}} = 1500$, bias-uncertainty ratios do not exceed $0.025$ in any case. In general, the systematic bias introduced by the spatially-varying $m$-bias on cosmological parameter estimates is expected to be sub-dominant, but not wholly negligible in certain cases. This is a refinement of previous work in this area, for example \cite{kitching:2019,kitching:2020} which found that the impact on the shear power spectra due to the spatially-varying $m$-bias should be small compared with the impact of the mean term; while we assume an $m$-bias with zero spatial mean and as such do not make comparative conclusions, we have quantified explicitly that the cosmological parameter bias induced by the spatially-varying $m$ may be a non-negligible contribution to the error budget. While the mean $m$ term may present the dominant contribution to the error budget from the $m$-bias, if this is well-constrained then the spatial variations may still result in a non-negligible bias in inferred cosmological parameter values. 
\subsection{Dependence of parameter bias on $m$-bias rms}
The amplitude of variations of the $m$-bias map which gives rise to parameter bias-uncertainty ratios  of $0.1, 0.3$ and $1$ is shown in figs. 7 and 8 for the 3-parameter and 4-parameter cases respectively, as a function of $\ell_{\mathrm{peak}}$, and for the optimistic high $\ell$-cut of $\ell_{\mathrm{max}} = 5000$. We also find that the dependence of bias-uncertainty ratio on $m$-bias rms is close to quadratic, with the exponent departing from 2 (negatively, by a few per cent) for $\ell_{\mathrm{peak}} \lesssim 500$. The quadratic relationship is is expected as the mode-coupling matrix is proportional to the $m$-bias angular power spectrum which scales as the variance of the $m$-bias map. The parameter bias-uncertainty ratio obtained from an $m$-bias map with a given rms depends on the scale of $m$-bias variations, and for an $m$-bias with a large characteristic scale the parameter bias can exceed the statistical uncertainty $\sigma$ for rms $\sim 0.04 - 0.05$; biases exceed $30 \%$ of the statistical uncertainty for rms $\sim 0.02 - 0.03$ for $m$-biases peaked across the range of scales examined. This shows that the importance of the spatially-varying $m$-bias is sensitive to the rms of the $m$-bias map, and hence that the potential impact of the spatially-varying $m$-bias on cosmological parameter constraints from cosmic shear cannot be neglected unless the expected amplitude of variations of the $m$-bias map is known from predicted constraints on the shear systematics which produce the spatially-varying $m$-bias. This allows requirements to be set on the allowed amplitude of variations of the spatially-varying $m$-bias, and hence on constraints of weak-lensing systematics from which the spatially-varying $m$-bias originates. In the case that the requirement on the amplitude of spatial variations of the $m$-bias field is satisfied, such that the resulting bias on inferred cosmological parameters is deemed to be negligible, then it will be possible to neglect the convolutive effect of the spatial variations of the $m$-bias on the shear power spectra, and consider the effect of the $m$-bias on the shear $C_{\ell}$s as a constant scaling arising due to the mean $m$. In this case it will be acceptable to mitigate the effect of the $m$-bias on cosmological parameter estimation by marginalising over a prior on the mean $m$, and the dominant effect of the $m$-bias on parameter inference will be the marginalisation effect considered by \cite{kitching:2020}.
\par
It should also be noted that while requirements exist on the value of the mean $m$-bias (the requirement from \cite{massey:2013} is that $m\lesssim2\times10^{-3}$) no such requirement has been placed on the amplitude of spatial variations of the $m$-bias (the rms of the $m$-bias field). In forthcoming surveys such as \textit{Euclid}, we will be interested in \textit{residual} shear biases, i.e. biases which result from uncertainties in our knowledge of the observational systematics that give rise to shear biases. Therefore, even if the mean $m$-bias is well calibrated subject to requirements, this does not by itself mean that the spatial variations of the residual $m$-bias must necessarily also be so; a key conclusion of this work is that it is separately necessary to place requirements on the amplitude of the spatial variations of the residual $m$-bias. This will allow requirements to be placed on models of the observational systematics that produce shear biases (eg. requirements on the precision of PSF models) and on priors of these systematics.
\section{Summary}
\label{summary}
We have considered the effect of a spatially-varying multiplicative shear bias on the estimation of cosmological parameters using the cosmic shear angular power spectrum. We have applied a computationally efficient pseudo-$C_{\ell}$ formalism to determine the bias in the cosmic shear power spectra $\smash{C_{\ell}^{\gamma\gamma}}$ arising due to an $m$-bias field obeying a $C_{\ell}$ with a Gaussian profile, by considering the $m$-bias map as acting as a mask that introduces mode mixing into the shear $C_{\ell}$s, and computing the mode-mixing matrix which convolves the $\smash{C_{\ell}^{\gamma\gamma}}$ with the power spectrum of the $m$-bias map. This has allowed us to consider the effect of the $m$-bias at high $\ell$, down to arcminute-scales corresponding to the expected field-of-view PSF variations in a Euclid-like photometric survey. We considered the settings for a stage-IV style lensing survey such as \textit{Euclid} \citep{euclidprep7:2020} and computed fiducial tomographic shear $C_{\ell}$s using \texttt{CCL} \citep{CCL:2019}. Note that while we consider \textit{Euclid} as an example, our indicative results are likely to be informative for other stage-IV weak-lensing surveys. We repeat this for a number of different $m$-bias maps characterised by different values for the mean and width of the $m$-bias map, $\ell_{\mathrm{peak}}$ and $\sigma_{m}$ respectively. We also calculate the $BB$ shear power spectra generated by the mode-coupling effect of the spatially-varying $m$-bias and finds that it follows the residual $EE$ shear power spectra closely, especially for a spatially-varying $m$-bias with a characteristic scale of arcminutes. We conclude that it may in principle be possible to use the induced $BB$ power to self-calibrate the spatially-varying $m$-bias, in agreement with previous work in this area (eg. \cite{kitching:2020}), though this will be complicated by the fact that the $BB$ power induced by different systematics may be degenerate, and so we do not consider this in our analysis.
\par
We find fractional residual $C_{\ell}$s which reach a maximum of $\sim 10^{-4}$. This is similar to the magnitude of fractional residuals found by \cite{kitching:2019,kitching:2020} between analytic calculations involving the spatially-varying $m$-bias and only the constant and mean terms up to a maximum multipole of 64, and between a numerical forward model including the spatially-varying $m$-bias and an analytic calculation involving only the constant and mean terms up to a maximum multipole of 2048, though we have extended the analytic calculation involving the spatially-varying $m$-bias to high $\ell$.
\par
We then employed a Fisher matrix analysis to forecast biases on cosmological parameters inferred from the $\smash{C_{\ell}^{\gamma\gamma}}$ due to the bias in the power spectra $\smash{\Delta C_{\ell}^{\gamma\gamma}}$ resulting from the spatially-varying $m$-bias, with no spatially constant term, i.e. assuming perfect calibration of the shear field for the spatially invariant $m$-bias. We find that the ratio of the parameter biases to the $1\sigma$ uncertainty depends strongly on $\ell_{\mathrm{max}}$ (a cut-off scale applied at the Fisher level) and $\ell_{\mathrm{peak}}$, with the parameter biases typically peaking both at high and low values of $\ell_{\mathrm{peak}}$, due to the dynamic nature of the $\smash{\Delta C_{\ell}^{\gamma\gamma}}$ as a function of $\ell_{\mathrm{peak}}$ and $\ell_{\mathrm{max}}$. We find that the parameter bias-uncertainty ratios are higher when more free parameters are included in the Fisher analysis. For realistic values of $\ell_{\mathrm{max}}$ for a \textit{Euclid}-style survey, and for $m$-bias map rms $= 0.01$, the bias in cosmological parameters resulting from a spatially-varying multiplicative shear bias reach a maximum of $\gtrsim 10\%$ of the forecast statistical error. Whereas \cite{kitching:2019,kitching:2020} concluded that the effect of a spatially-varying m-bias should be small compared with that of the mean $m$-bias, and negligible in the case of an $m$-bias field with zero mean, we find that the effect of the spatially-varying m-bias can be neglected only subject to requirements on the properties of the $m$-bias field, in particular the rms and characteristic scale of spatial variations. This caveat arises because we consider the impact of the $\smash{\Delta C_{\ell}^{\gamma\gamma}}$ up to significantly higher multipoles, which allows us to explicitly calculate the effect on the inference of cosmological parameters. This requires that the spatially-varying $m$-bias be a spin-0 field, though this is known to be a realistic assumption (for example, \cite{kitching:2019} found that the effect on the shear power spectra due to the imaginary part of the $m$-bias should be very small compared with the real part). This simplification allows us to make use of an orthogonality relation in the computation of the $m$-bias mode-coupling matrix which significantly reduces the scaling of the calculation with the maximum multipole considered compared with the formalism of \cite{kitching:2019}.
\par
We have also investigated the variation of the parameter biases with the rms of the $m$-bias map, and find that the relationship is close to quadratic. We find that biases exceed $30\%$ of the statistical error for rms $\sim 0.02 - 0.03$ across the range of $\ell_{\mathrm{peak}}$ examined, and exceed the statistical error for rms $\sim 0.04-0.05$ for small $\ell_{\mathrm{peak}}$. This allows requirements to be set on the permissible amplitude of variations of the $m$-bias, and hence on models of the relevant shear systematic effects that will be used for systematics control in forthcoming surveys. If the spatial variations of the $m$-bias field satisfy such a requirement, then it may be possible to mitigate the effect of the $m$-bias on parameter inference by marginalisation, as considered by \cite{kitching:2020}.
\par
In this work, we have considered a simple model for spatially-varying $m$-bias, in which the $m$-bias field exhibits power on a specific scale. This allows us to consider the sensitivity of inferred cosmological parameters to generic $m$-bias, arising due to systematics with power on a specific physical scale. However, a more realistic model may consider the potential redshift dependence of the $m$-bias, or a more realistic prescription for the statistics of the $m$-bias field, including the $m$-bias $C_{\ell}$. In particular, it will be of interest to consider the effect of specific systematics which may contribute to the shear bias, including PSF and other instrumental effects, as well as effects relating to target selection, galaxy shape measurement, etc. In considering such effects it may also be of interest to consider the effect of the additive shear bias in a related analysis, which may potentially be considered in the pseudo-$C_{\ell}$ framework.
\section*{Acknowledgements}
We thank T. D. Kitching for useful discussions. CC is supported by an STFC doctoral studentship. CAJD acknowledges support from the Beecroft trust. DA acknowledges support from the Beecroft trust and from STFC through an Ernest Rutherford Fellowship, grant reference ST/P004474/1. We are thankful to the authors of \texttt{NaMaster}, \texttt{CCL} and \texttt{CLASS} for developing and making their codes publicly available.
\FloatBarrier
\bibliographystyle{mnras}
\bibliography{biblio}

\bsp	
\label{lastpage}
\end{document}